\documentclass{myczjphys}         
\RequirePackage{epsfig}

\begin{document}
\title{Non-Hermitian Quantum Theory and its Holomorphic Representation: Introduction and Some Applications}  
%
\authori{Frieder Kleefeld\,\footnote{E-mail: {\sf kleefeld@cfif.ist.utl.pt}\,, URL: {\sf http:/$\!$/cfif.ist.utl.pt/$\sim$kleefeld/}}}      \addressi{Centro de F\'{\i}sica das Interac\c{c}\~{o}es Fundamentais (CFIF), Instituto Superior T\'{e}cnico,\\ Av.\ Rovisco Pais, 1049-001 Lisboa, Portugal}
\authorii{}     \addressii{}
\authoriii{}    \addressiii{}
\authoriv{}     \addressiv{}
\authorv{}      \addressv{}
\authorvi{}     \addressvi{}
%
\headauthor{Frieder Kleefeld}            
\headtitle{Non-Hermitian Quantum Theory and its Holomorphic Representation \ldots}             
\lastevenhead{Frieder Kleefeld: Non-Hermitian Quantum Theory and its Holomorphic Representation \ldots} 
\refnum{A}
\daterec{XXX}    
\issuenumber{?}  \year{2004}
\setcounter{page}{1}
\maketitle

\begin{abstract} Present Hermitian Quantum Theory (HQT), i.e.\ Quantum Mechanics (QM) and Quantum Field Theory (QFT), is revised and replaced by a consistent non-Hermitian formalism called non-Hermitian Quantum Theory (NHQT) or (Anti)Causal Quantum Theory ((A)CQT) after lining out some inherent inconsistencies and problems arising in the context of causality, which is observed to introduce an indefinite metric in canonical commutation relations. Choosing some (very selective) historical approach to introduce necessary terminology and explain complications when quantizing non-Hermitian systems in the presence of an indefinite metric we propose a way how to construct a causal, analytic, Poincar\'{e} invariant, and local NHQT, the spacial representation of which is related to the so-called holomorphic representation used in complex analysis. Besides providing a revised antiparticle, spinor, and probability concept, a new neutrino Lagrangean, two distinct time-reversal operations, and generalized non-Hermitian Poincar\'{e} transformations, we will apply NHQT to consider three important issues: PT-symmetry and non-Hermitian similarity transforms, non-Hermitian supersymmetry, and the construction of an asymptotically free theory of strong interactions without gluons.
\end{abstract}

\section{Introduction: Some History, Terminology, Consistency Arguments}
During his remarkable Bakerian Lecture at June 19, 1941, the published version of which dates back to 1942 \cite{Dirac:1942}, P.A.M.\ Dirac summarized in his decisive manner a by now very influential conception of the physical world to which he had contributed more than significantly (See also e.g.\ \cite{Quinn:2001kb,Dalitz:1995,heis1972}!) \footnote{According to Dirac Fermionic particles are not only occupying some of the positive energy eigenstates of a system under consideration, yet almost all its negative energy eigenstates (``Dirac sea'') \cite{Dirac:1930}. Beyond eventually existing unoccupied states with postitive energy in a Fermion solid (``holes''), feasible unoccupied negative energy states in the Dirac sea are identified with Fermionic antiparticles. The Fermionic vacuum is at $E\rightarrow - \infty$. As Bosonic particles do not follow Pauli's exclusion principle, Dirac's concept of filling up negative energy states could not be applied to Bosonic systems. Hence, Dirac positioned the Bosonic vacuum at $E=0$, in order to avoid Bose-Einstein condensation of Bosonic particles towards $E\rightarrow - \infty$. Admitting hereby Bosonic particles of only positive energy Dirac found nevertheless a very elegant way to make use of the negative energy eigenstates of a Bosonic system. Dirac noticed that Bosonic states of negative energy have simultaneously also {\em negative norm} and {\em negative probability} (See also \cite{Pauli:1943,mueckenheim1}!). Therefore Dirac \cite{Dirac:1942,Dirac:1943} declared Bosonic negative energy states to be {\em responsible for the annihilation (i.e.\ absorption) of positive energy particles} (The negative energy states discussed in \cite{Henry-Couannier:2004mn} reflect and support nicely this interpretation of Dirac (even for Fermions), even though completely differently interpreted by the respective author!). Strictly speaking, Dirac associated even Bosonic positive energy states with the {\em emission} of Bosonic particles and {\em not} with their {\em existence}!}. 

From a modern point of view the conception of Dirac has besides a rather asymmetric treatment of Fermions and Bosons its particular drawbacks in two points causing serious difficulties in their theoretical handling: the mere existence of a ``Dirac sea'' with its infinite negative energy (causing in particular problems to cosmologists) and the missing description of Bosonic antiparticles \footnote{To our understanding the great merit of Dirac's work superseding by far its drawbacks lies --- ironically --- in the fact that Dirac provides us in \cite{Dirac:1942} simultaneously with a great collection of theoretical tools allowing us to remove the drawbacks in his conception of the physical world!}.

In Vol.\ 1 of his book ``The Quantum Theory of Fields'' \cite{Weinberg:1995mt} S.\ Weinberg seems to suggest that both problems have been surpassed by formalisms developed 1934 by W.\ Pauli and V.\ Weisskopf (for spin 0 Bosons) \cite{Pauli:1934xm} and W.H.\ Furry and J.R.\ Oppenheimer (for spin 1/2 Fermions) \cite{Furry:1934} \footnote{On p.\ 23 ff in \cite{Weinberg:1995mt} he states with respect to W.H.\ Furry and J.R.\ Oppenheimer: 
{\em ``$\ldots$ Furry and Oppenheimer picked up Dirac's idea that the positron is the absence of a negative-energy electron; the anticommutation relations are symmetric between creation and annihilation operators, so they defined the positron creation and annihilation operators as the corresponding annihilation and creation operators for negative-energy electrons  $b^+_k \equiv a_k$, $b_k\equiv a^+_k$ (for $\omega_k <0$) where the label $k$ on $b$ denotes a positive-energy positron mode with momenta and spin opposite to those of the electron mode $k$. $\ldots$ it is necessary also to specify that the physical vacuum is a state $\Psi_0$ containing no positive-energy electrons or positrons: $a_k \Psi_0 = 0$ ($\omega_k>0$), $b_k\Psi_0 = 0$ ($\omega_k<0$) $\ldots$''}.
Then, in the context of the work of W.\ Pauli and V.\ Weisskopf, he remarks on p.\ 26 ff in \cite{Weinberg:1995mt}:
{\em ``$\ldots$ The existence of two different kinds of operators $a$ and $b$, which appear in precisely the same way in the Hamiltonian, shows that this is a theory with two kinds of particles with the same mass. As emphasized by Pauli and Weisskopf, these two varieties can be identified as particles and the corresponding antiparticles, and if charged have opposite charges. Thus as we stressed above, bosons of spin zero as well as fermions of spin 1/2 can have distinct antiparticles, which for bosons cannot be identified as holes in a sea of negative energy particles  $\ldots$''}.
In the course of time even Dirac changed --- according to S.\ Weinberg --- his point of view as is stated on p.\ 13 ff in \cite{Weinberg:1995mt}: {\em ``$\ldots$ And if the hole theory does not work for bosonic antiparticles, why should we believe it for fermions? I asked Dirac in 1972 how he then felt about this point; he told me that he did not regard bosons like the pion or $W^\pm$ as `important'. In a lecture a few years later, Dirac referred to the fact that for bosons `we no longer have the picture of a vacuum with negative energy states filled up', and remarked that in this case `the whole theory becomes more complicated'. $\ldots$''}.}. {\em If} one characterizes stable particles serving as asymptotic states by {\em real} energies, then one might be tempted to agree {\em for such real energy particles} to the view of Pauli, Weisskopf, Furry and Oppenheimer (PWFO) (adverted by Weinberg and finally also by Dirac), as the introduction of two operators $a_k$ and $b_k$ for particles and antiparticles, respectively, being related by $b^+_k \equiv a_k$ and $b_k\equiv a^+_k$ for $\omega_k <0$ indeed admits the vacuum to be positioned at $E=0$, while the operator $b_k$ reflects even without a Dirac sea all properties of holes in Dirac's sea of negative {\em real} energy states. 
Furthermore the concept has the advantage to be applicable not only to Fermions, but also to Bosons \footnote{Even the ``proof'' of negative relative intrinsic parity between Fermions and anti-Fermions by V.B.\ Berestetskii  seems to remain in force \cite{Weinberg:1995mt}. Several people (including the editor of \cite{shi2001a}  on p.\ 53) claim that the ``proof'' of the negative relative sign between the intrinsic parity of a Fermionic particle and its antiparticle has been performed for the first time by V.B.\ Berestetskii \cite{ber1980a}.}. Yet --- as will be explained in more detail below --- the view of PWFO {\em cannot} be kept as it stands without inconsistency, when one tries to formulate a local, causal, analytic and Poincar\'{e}-covariant Quantum Theory (QT) for systems with complex-valued selfenergies and vertex-functions \footnote{This exercise is guided by the observation that besides the few stable particles which may serve as asymptotic states the majority of (anti)particle-like excitations in nature are short-lived and resonance-like.}. The result of such an exercise \cite{Kleefeld:2003dx,Kleefeld:2003zj,Kleefeld:2002au,Kleefeld:2002gw,Kleefeld:2001xd,Kleefeld:1998yj,Kleefeld:1998dg,Kleefeld:thesis1999} may be called NHQT or (A)CQT and will be described to some extend in the following text. It will yield a ``revised'' conception of the physical world treating Fermions and Bosons completely symmetrically and being free of the above conjectured inconsistency arising not only in the early conception by Dirac, but also in the subsequent, more sophisticated conception by PWFO adverted e.g.\ by Weinberg (See also \cite{Henry-Couannier:2004mn}!). This inconsistency, the resolution of which is a striking argument in favour of (A)CQT, may be sketched as simple as follows:

Assume that $a_k(\omega_k)$ is the annihilation operator of a {\em causal} particle with {\em complex} energy $\omega_k$ (Re$[\omega_k]> 0$, Im$[\omega_k]< 0$) \footnote{Massless particles are to be considered by a limiting procedure preserving their causal features.}. Then the respective creation operator of the particle with {\em complex} energy $\omega_k$ is $a_k(-\omega_k)$ ($\equiv b^+_k$ according to PWFO). The Hermitian conjugation of $a_k(-\omega_k)$ leads then to $a^+_k(-\omega_k)$ ($\equiv b_k$ according to PWFO) being the annihilation operator of an {\em anticausal} particle with the {\em complex conjugated} energy $\omega_k^\ast$, and not to the annihilation operator of a {\em causal} antiparticle with energy $\omega_k$, as suggested by PWFO. For completeness we note that $a^+_k(\omega_k)$ is the creation operator of the {\em anticausal} particle with energy $\omega_k^\ast$ \footnote{As causality is an important issue not only for short-lived intermediate excitations in scattering processes, yet also for very long-lived particles serving as asymptotic states (displaying their causal behaviour by an infinitesimal negative imaginary part in their mass), we have to realize that the previous argument {\em ruling out} the approach to antiparticles by PWFO applies also to causal particles or antiparticles with a seemingly real mass like the electron or positron, respectively.} \footnote{It is interesting to see how even the authors of \cite{machet2004} gloss over these causality properties of the considered $K^0$ and $\bar{K}^0$ mesons when performing the identification $\phi_{\bar{K}^0}(\vec{x})=\exp(-i\delta)\,\phi^+_{K^0}(\vec{x})$.}. 

This unexpected lesson from a non-Hermitian causal approach to QT should not surprise, yet be seriously taken into account, as illustrated also by the following example. As early as 1954 T.D.\ Lee observed in the so-called {\em normal} ``Lee-model''~\cite{Lee:1954iq} (See also e.g.\ \cite{kaellen1955,Pauli:1956,glaser1956,heis1957,heis1961,nakanishi1958,ascoli1959,ferretti1959,nagy1960a,nagy1960b,sudarshan1961a,sudarshan1961b,vaughn1965,nagy1966,Lee:1969fz,Kleefeld:Nakanishi:1972pt,hittner1973,casagrande1973,Bender:1974pv,nagy1974,Rabuffo:1977va}!) being a somehow extended version of the preceding ``Friedrichs-model'' \cite{Friedrichs:1948} of 1948 a similar surprise due to unexpected non-Hermiticities, which he comments as follows: {\em ``$\ldots$ it can be shown that the result of the renormalization process cannot be obtained by any limiting process that involves only real values of the unrenormalized coupling constant. This difficulty may, however, be overcome by a modification of the present rules of quantum mechanics. $\ldots$''} \cite{Lee:1954iq} \footnote{Recall that the (unrenormalized) {\em normal} $(+)$ \cite{Lee:1954iq} and {\em abnormal} $(-)$ \cite{kaellen1955} Lee-model described according to \cite{heis1957,heis1961,Rabuffo:1977va} by a Hamilton operator $H = H_0 + H_{int}$ with $H_0 = \pm m_V \, \psi^+_V \psi_V + m_N \, \psi^+_N \psi_N + \int d^3k \; a^+(\vec{k}) \; a(\vec{k}) \, \omega (\vec{k})$ and
$H_{int} = - \, \frac{\displaystyle g_0}{\sqrt{4\,\pi}} \int^{\hat{\omega}} \! d^3k \; \frac{f(\omega(\vec{k}))}{(2\, \omega(\vec{k}))^{1/2}} \big[ \,\pm \psi^+_V \, \psi_N \;  a(\vec{k}) \; + \; a^+(\vec{k}) \; \psi^+_N \, \psi_V \, \big]$ consists of a {\em normal/abnormal} Fermion $V$ and a {\em normal} Fermion $N$ with no kinetic energy coupling to the {\em positive frequency part} $\varphi^{(+)}$ of a Boson $\varphi$ (``$\theta$-particle''), i.e. $V \rightleftharpoons N + \varphi^{(+)}$!}. 
A subsequent discussion by W.~Pauli~\cite{Pauli:1956} in June, 1955, based on a joint analysis with G.\ K\"all\'{e}n \cite{kaellen1955} resulting in the so-called {\em abnormal} Lee-model clarified in a very decisive and instructive way, what is going on in the Lee-model: the Lee-model's coupling constant renormalization $N^2\equiv g^2/g^2_0 = 1/(1+A\,g^2_0)$ with $A>0$ determined by the square of the renormalized coupling $g$ and the unrenormalized coupling $g_0$ changes sign when violating the inequality $0<g^2<1/A$ yielding --- as already observed by Lee --- a {\em purely imaginary} unrenormalized coupling $g_0$ for a  finite, {\em purely real} renormalized coupling $g$. 
While Lee \cite{Lee:1954iq} was interpreting the unexpected change of the Hermiticity character between the {\em non-Hermitian unrenormalized} Hamilton operator resulting from a {\em purely imaginary} coupling $g_0$ and the {\em Hermitian renormalized} Hamilton operator resulting from a {\em real} coupling $g$ as a {\em non-unitary similarity transform} \footnote{According to R.F.\ Streater \cite{streater2004} one calls a (not necessarily selfadjoint) observable similar to a real diagonal operator a {\em diagon}.}, Pauli \cite{Pauli:1956} stressed that the violation of the inequality $0<g^2<1/A$ {\em ``$\ldots$ leads to a contradiction with the concept of physical probability (indefinite metric of the Hilbert space) $\ldots$ connected with the appearance of new discrete stationary states whose contribution to the conserved sum of `probabilities' is negative (`ghosts') $\ldots$''} \cite{Pauli:1956}~\footnote{To our best knowledge (See also p.\ 509 in \cite{enz2002}!) this is the place where the term {\em ghost} enters --- in the form of a definition by W.\ Pauli --- for the first time physics literature.}. 
According to foregoing literature Pauli called states with positive norm (and probability) {\em normal} and states with negative norm (and probability) {\em abnormal} \footnote{Abnormal Fermions are addressed e.g.\ by Pauli \cite{kaellen1955,Pauli:1956} (e.g.\ in the abnormal Lee-model) and Lee \& Wick \cite{Lee:iw}. Hence, they are not invented in \cite{Mostafazadeh:2004ph}, where they are called {\em abnormal Phermions}!} (See also \cite{Pauli:1943,nagy1960b,nagy1966}!). 
Pauli's and K\"all\'{e}n's observation about the existence of an indefinite metric and abnormal states in the Lee-model allowed them not only to understand that the Hamilton operator of their {\em abnormal} Lee-model is {\em pseudo-Hermitian} \footnote{The meaning of the terminology {\em pseudo-Hermiticity} frequently used by W.\ Heisenberg \cite{heis1961,nagy1960a} and {\em pseudo-unitarity} used already 1958 by W.\ Pauli \cite{pauli1958} on the basis of an {\em indefinite metric} was in the same year discussed e.g.\ by L.K.\ Pandit \cite{pandit1959}. Hesitating to distinguish between {\em unitarity} (being reserved for the transformations which preserve the usual positive definite metric) and {\em pseudo-unitarity} (i.e.\ unitarity with respect to a specific metric) he truely stated: {\em ``$\ldots$ The concept of unitarity cannot be defined without specifying the metric$\ldots$''} (For subtile critics of this understanding from the point of view of an afterwards introduced {\em new} terminology see e.g.\ \cite{Mostafazadeh:2003iz}!).} with respect to this indefinite metric, yet also, what is actually going on in the subspace ($V,N+\theta$) of the Lee-model during the process of violating the inequality $0<g^2<1/A$ by increasing~$g^2$~\cite{ascoli1959}~\footnote{The real energy eigenvalue of the {\em normal} eigenstate of the $V$-particle collides at a critical value of $g$, when $0<g^2<1/A$ is violated and $N^2=g^2/g^2_0$ vanishes, with a real energy eigenvalue of an additional {\em abnormal} eigenstate; this collision point (or {\em critical point}) of the abnormal Lee-model is an {\em exceptional point}, as the eigenspace of the Hamilton operator spanned by the normal and the abnormal eigenstate reduces at the exceptional point its dimension by one and is now characterized by one state only called {\em good ghost} \cite{duerr1969,heis1972} due to its zero norm resulting from a now degenerate metric; fortunately there can be constructed an additional state called {\em dipole ghost} \cite{Pauli:1956,heis1957,heis1961} or {\em bad ghost} \cite{duerr1969,heis1972}, if chosen such (See p.\ 111 in \cite{nagy1960b}!) that its norm vanishes (In consequence the good and bad ghost form a {\em biorthogonal basis}!); this dipole ghost is required to preserve the original dimension of the space spanned by the normal and abnormal state before their collision; a further increase of $g^2$ beyond the critical value leads then to the formation of a pair of --- again --- {\em biorthogonal states} (called e.g.\ ``complex ghosts'' \cite{Kleefeld:Nakanishi:1972pt}, ``complex roots'' \cite{pauli1958}, ``complex poles'' \cite{lee1970}, or ``exponential ghosts'' \cite{froi1959}) of {\em zero (traditional) norm} with complex valued, mutually complex conjugate eigenvalues (Note that such a bifurcation from real to complex pairs of energy eigenvalues does not only occur in the Lee-model. It takes place e.g.\ in Quantum Electrodynamics (QED) at the Landau-pole and in Quantum Chromodynamics (QCD) at the confinement-deconfinement phase-transition!). The collision scenario  in the Lee-model involving just a dipole ghost is just a special case of the more general situation, when there develop higher order degeneracies with non-trivial Jordan-Block structure \cite{wilk1965,Puntmann:1997nm,Bohm:1997mr} at the critical point, i.e.\ so-called {\em multipole ghosts} \cite{pandit1959,nagy1960a,nagy1960b,nagy1966,Kleefeld:Nakanishi:1972pt}. The theory of parameter dependent flow, degeneracy and bifurcation of eigenvalues at or around critical points has made --- since the work of A.M.\ Lyapunov in 1892 \cite{lyapunov1892} (See also \cite{argyris1994}!) --- in several fields of theoretical physics (chaos theory, renormalization group theory, $\ldots$) great progress. Relevant references also in the context of exceptional/diabolic points may be found in the work of M.V.\ Berry \cite{berry2004} and other research groups, e.g.\ \cite{okolowicz2004,Heiss:1998bv,Puntmann:1997nm,Bohm:1997mr}.}. 
The analysis of the surprising properties of the Lee-model required and hence induced  --- similarly as the analysis of the antiparticle concept discussed above --- not only a much deeper understanding of formalistic aspects of QT --- in particular in the presence of an {\em indefinite metric} \footnote{The development of a formalism for a QT with an indefinite metric being initiated most probably by P.A.M.\ Dirac \cite{Dirac:1943,Dirac:1942} and W.\ Pauli \cite{Pauli:1943} in 1942/1943 has made since then great progress \cite{heis1972}. Selective subsequent very early and influential related physics publications are certainly \cite{Pauli:1949zm,Gupta:1950,Bleuler:1950cy,kaellen1955,Pauli:1956,heis1957,gupta1957,ascoli1958b,konisi1958,konisi1959,froi1959,sudarshan1961a,sudarshan1961b,schlieder1961,heis1961,gazdy1980} (See also most references mentioned in the context of the Lee-model!). For some important reviews written by physicists including a lot of important references we refer to L.K.\ Pandit (1959 \cite{pandit1959}), K.L.\ Nagy (1960 \cite{nagy1960b}, 1966 \cite{nagy1966}, 1974 \cite{nagy1974}), and N.\ Nakanishi (1972 \cite{Kleefeld:Nakanishi:1972pt}). The ``$\eta$-formalism'' (presumably) by W.\ Pauli \cite{Pauli:1943} ($\eta \equiv$ ``metric operator''), which was already thoroughly discussed e.g.\ in \cite{pandit1959,nagy1960b,nagy1966}, has recently been intensively ``rediscovered'' by A.\ Mostafazadeh (See e.g.\ \cite{Mostafazadeh:2003iz} and references therein!). It is interesting to observe that the mathematical literature on operators in spaces with indefinite metric seems to start approximately 1944, i.e.\ shortly after P.A.M.\ Dirac and W.\ Pauli, by contributions of L.S.~Pontrjagin (1944) \cite{pontrjagin1944}, M.G.\ Kre\u{\i}n ($\simeq$ 1950) \cite{krein1950,iohvidov1956}, R.\ Nevanlinna (1952,1956) \cite{nevalinna1952}, and I.S.\ Iohvidov (1956,1959) \cite{iohvidov1956} (See also references therein!). Selective early mathematical works suitable for consideration by physicists might be also \cite{louhivaara1958,scheibe1960,langer1962}. In the meantime there have appeared several well written mathematical reviews on the huge amount of related mathematical literature, e.g.\ \cite{bognar1974,iohvidov1982,azizov1989,dritschel1996,langer1998}, while relevant contributions of mathematical physicists (e.g.\ \cite{Jakobczyk:1983ck,Jakobczyk:1986gx,Broadbridge:1982xq}) seem to have developed their own characteristic language. Following \cite{dritschel1996} we want here just to recall some important terminology: an {\em inner product space} is a complex vector space ${\cal V}$ together with a complex valued function $\left<\cdot,\cdot\right>=\left<\cdot,\cdot\right>_{\cal V}$ ($\equiv$ {\em inner product}) on ${\cal V} \times {\cal V}$ which satisfies the axioms of {\em linearity} ($\left<f,a \,g + b \,h\right> =a \, \left<f, g \right> + b \, \left<f,h\right>$ for all $f,g,h\in{\cal V}$ and $a,b\in {\bf C}$) and {\em symmetry} ($\left<f,g\right>=\left<g,f\right>^\ast$ for all $f,g\in{\cal V}$); the {\em antispace} of an inner product space $({\cal V},\left<\cdot,\cdot\right>)$ is the inner product space  $({\cal V},- \left<\cdot,\cdot\right>)$; a {\em Hilbert space} is a strictly positive inner product space ${\cal H}$ over the complex numbers which is complete in its norm metric; a {\em Kre\u{\i}n space} ${\cal K}$ is an orthogonal direct sum of an Hilbert space ${\cal H}_+$ and its antispace ${\cal H}_-$, i.e.\ ${\cal K}= {\cal H}_+ \oplus {\cal H}_-$; a {\em Pontrjagin space} ${\cal P}$ is a Kre\u{\i}n space with a finite negative index, i.e.\ $\mbox{ind}_-\,{\cal P} <\infty$; note that $\mbox{ind}_\pm \,{\cal K} =\mbox{dim} \,{\cal K}_\pm$.} and {\em complex energy eigenvalues} (See e.g.\ \cite{okolowicz2004,gazdy1980,gadella2004,moiseyev1998,Hagen:2003ev}!) ---, it raised also serious new questions. One immediately arising question motivated by W.\ Heisenberg \footnote{According to p.\ 412 in \cite{duerr1969} {\em ``$\ldots$ it was mainly Heisenberg who emphasized that the existence of a unitary S-matrix for physical states will be sufficient to guarantee the usual quantum-mechanical probability interpretation. $\ldots$''}}, i.e.\ whether the scattering matrix of a theory containing simultaneously (complex) ghosts and eigenstates of strictly real energy eigenvalues can be unitary, was very early answered positively by W.~Pauli~\cite{pauli1958} (See also \cite{ascoli1958}!) and W.\ Heisenberg \cite{heis1957} (See also \cite{duerr1970}!). This result was later confirmed by T.D.\ Lee \& G.C.\ Wick \cite{Lee:1969fz,lee1970,Lee:1969fy} and H.P.~D\"urr \& E.~Seiler \cite{duerr1970} (See also \cite{ferretti1959,Broadbridge:1982xq,demuth1981}!). 
As will be sketched below, another serious and unresolved question remained until very recently, how to achieve beyond unitary simultaneously also {\em analyticity}, {\em causality}, {\em locality} and {\em Poincar\'{e}} or {\em Lorentz invariance} in theories containing (complex) ghosts. As in the context of the antiparticle concept discussed above this puzzling point can be resolved within the new framework of (A)CQT. What concerns the issue {\em analyticity}, the situation had been summarized along to the results obtained in \cite{Gleeson:1972xj,duerr1970} on p.\ 135 in \cite{heis1972} by W.\ Heisenberg: {\em ``$\ldots$ Taking the dipole case as an example it is easily seen that the amplitudes (or related functions) cannot be analytic at a possible threshold to the creation of ghost-states, because the boundary conditions must be different below and above threshold.$\ldots$''}. According to \cite{Gleeson:1972xj} (See also \cite{Sudarshan:1974}!) he calls such an unsatisfactory situation {\em piecewise analyticity}. 
The complications with {\em causality}, {\em locality} and {\em Lorentz invariance} were approached 1968 in an instructive manner by E.C.G.\ Sudarshan et al.\ \cite{Arons:1969gp,Dhar:1969gb} in considering the field theory of a {\em tachyonic} Klein-Gordon (KG) field with {\em purely imaginary mass} \footnote{This investigation finds its motivation in E.P.\ Wigner's observation that the existing irreducible representations of the Poincar\'{e} group factorize into 6 cases \cite{wigner1963}: 1) $p_\mu \,p^\mu =m^2$, $p^0>0$; 2) $p_\mu \,p^\mu=0$, $p^0>0$; 3) $p_\mu \,p^\mu<m^2$; 4) $p_\mu \,p^\mu =0$, $p^0<0$; 5) $p_\mu \,p^\mu=m^2$, $p^0<0$; 6)~$p^0=p^1=p^2=p^3=0$. Case 3) discussed on p.\ 76 in \cite{wigner1963} is obviously tachyonic!}. They observed that in such a seemingly Poincar\'{e} invariant theory \footnote{Statements of Sudarshan et al.\ have to be taken with great precaution. O.W.\ Greenberg \cite{Greenberg:2004vt} showed how Lorentz-covariance of time-ordered products implies microcausality, i.e. spacelike local (anti)commutativity. Hence, nonlocal theories are in general expected to be {\em not} Lorentz-covariant.} with {\em real} energies and 3-momenta on one hand positive (negative) energy states can be transferred to negative (positive) energy states, respectively, by Lorentz transformations, on the other hand that a consistent construction of interaction terms involves causal {\em and} anticausal propagators simultaneously yielding {\em non-local interactions} and a {\em causality violating} theory. Furthermore they note \cite{Dhar:1969gb}: {\em ``$\ldots$ Any attempt at avoiding Fock states for negative-energy particles violates the relativistic invariance of the theory. $\ldots$''}.   It is surprising to observe that already in 1959 M.\ Froissart \cite{froi1959} considered a --- to our present knowledge --- much more well behaved and Lorentz covariant model of two Hermitian conjugate complex ghosts with complex mass $M$ described by the Lagrange density ${\cal L} (x) =\frac{\alpha}{2} \left( (\partial \phi (x) )^2  - M^2 \, \phi (x)^2 \right) + 
\frac{\alpha^\ast}{2} \left( (\partial \phi^+ (x) )^2  - M^{\ast \, 2} \, \phi^+ (x)^2 \right)$ with \footnote{Even though the chosen phase $\alpha=-i$ allows quite nicely to illustrate the concept of canonical quantization in the presence of a dipole ghost within a relativistic context and related complications in the formulation of a S-matrix, it simultaneously yields a theory lacking physical significance for the description of nature.} $\alpha=-i$. In choosing --- without loss of generality --- Re$[M]>0$ and Im$[M]<0$ we will call in the follwoing the fields $\phi(x)$ {\em causal} and the fields $(\phi(x))^+$ {\em anticausal}. 13 years later, in 1972, N.\ Nakanishi \cite{Kleefeld:Nakanishi:wx,Kleefeld:Nakanishi:1972pt} then studied Froissart's Lagrange density with the more significant phase choice $\alpha=1$ under the name {\em Complex-Ghost Relativistic Field Theory} not refering to Froissart in his first submitted work \cite{Kleefeld:Nakanishi:wx}, yet to work by T.D.\ Lee \& G.C.\ Wick \cite{Lee:iw} of 1970 who considered a similar problem ironically already for the more complicated case of vector and Dirac fields with complex mass (in the presence of a {\em purely imaginary} gauge-coupling constant!) \footnote{It is interesting to note that also Lee \& Wick did not refer in \cite{Lee:iw} to the work of Froissart.}. As a result of an intensive analysis of the model he emphazised on p.\ 68 in \cite{Kleefeld:Nakanishi:1972pt} that {\em ``$\ldots$ the relativistic complex-ghost field theory is a non-trivial, divergence-free quantum field theory which is manifestly covariant at any finite time and whose physical S-matrix is unitary and macrocausal. So far we know no other theory which has all these features. The Lorentz non-invariance of the physical S-matrix is not necessarily its demerit but it can be a merit. $\ldots$''}. 
Obviously N.~Nakanishi concluded his analysis of the theory --- apart from an outline of its many beautiful features reveiled --- with a general, discouraging claim about its Lorentz non-invariance \footnote{It is important to note that N.\ Nakanishi repeats here an argument about Lorentz non-invariance, which he had made \cite{Nakanishi:1971jj,Kleefeld:Nakanishi:wx} (See also \cite{Lee:1971ix,Kleefeld:Nakanishi:1972pt}!) already 1970 in the context of the mentioned intimately related model by Lee \& Wick \cite{Lee:iw} of the same year. It consists essentially of the observation that possible loop-diagrams containing a pair of complex ghosts $\phi(x)$ and $(\phi(x))^+$ with complex mass $M$ and $M^\ast$, respectively, which were in great detail discussed by A.M.\ Gleeson et al.\ \cite{Gleeson:1972xj} at the end of 1970 and led there to the problem of {\em piecewise analyticity} mentioned above, are not Lorentz-invariant.}. Hence, even being not tachyonic Nakanishi's Complex-Gost Relativistic Field Theory seemed to suffer from similar problems as the tachynonic KG theory considered by Sudarshan et~al.\ in 1968. 
Yet, contrary to the tachynonic KG theory studied by Sudarshan et al., there is fortunately --- as conjectured by the author --- a very simple way to cure the lack of Lorentz-invariance, analyticity, causality and locality in the Complex-Ghost Relativistic Field Theory of N.\ Nakanishi: it is our POSTULATE that there should be {\em no interaction terms between causal fields $\phi(x)$ and anticausal fields $(\phi(x))^+$ in the Lagrange density} \footnote{This postulate leads to the {\em absence} of loops containing {\em both} kind of fields and, in turn, requires that even asymptotic states have to be treated like complex ghosts with complex-valued mass containing an infinitesimal imaginary part, which is quite consistent with all the features of in- and out-states in a non-stationary description of standard Hermitian QT (HQT). Note that asymptotic states of strictly real mass (or energy) would be a superposition of causal and anticausal states, which would typically couple to intermediate states such that they violate our postulate!}. 
The resulting QT called (A)CQT being described in the following text has the nice feature to be Lorentz covariant, analytic, causal and local even in the presence of fields of complex mass with finite imaginary part describing short-lived intermediate particles, which is due to a {\em very tricky interplay} between the underlying normal and abnormal states having caused so many troubles and puzzles since the Lee-model. In summary --- contrary to HQT --- in (A)CQT the causal and the anticausal sector of the theory are completely disconnected, which is the price we have to pay to achieve simultaneously Lorentz covariance, causality, locality and analyticity \footnote{Making use of the whole complex plane we additionally have to take into account in (A)CQT --- as will be clarified in the following text --- a {\em generalized probability} concept, which allows not only real positive and negative probabilities \cite{mueckenheim1}, yet in general {\em complex probabilities} \cite{berggren1970}. It is just the privilege of causal or anticausal asymptotic in- and out-states to be characterized by probabilities being infinitesimally close to real numbers.}.

It is interesting to observe present strong renewed interest (See e.g.\ \cite{pseudo2003}!) in quantum-theoretic questions of the Lee-model by physicists and mathematicians, yet now under a different headline called {\em PT-symmetry} \cite{Bender:1998ke,Znojil:2001,Znojil:2004xw}, i.e.\ symmetry under space- and time-reversal \footnote{The field rooted 1980 in the observation \cite{Caliceti:1980} (See also \cite{Delabaere:1998}!) that some part of the spectrum of non-Hermitian Hamilton operators like $H=p^2+x^2+i\,x^3$ may be real. Based on a fascinating conjecture by D.\ Bessis (and J.~Zinn-Justin) of 1992, that the complete spectrum of this Hamilton operator is real and positive, C.M.\ Bender and S.\ Boettcher \cite{Bender:1998ke} (See also \cite{Bender:1998gh}!) suggested in 1997 that a whole class of such Hamilton operators possess this feature due to their {\em antiunitary} \cite{robnik1986} {\em PT-symmetry}. 
These developments were accompanied by related investigations in the context of (anharmonic) quartic oscillators (See e.g.\ \cite{buslaev1993,Fernandez:1998}). A review provided in 2001 by M.\ Znojil as a preprint {\sf math-ph/0104012} (unfortunately published very delayed in 2004 \cite{Znojil:2001}) containing a lot of important references to related work showed that the new research field had again, yet independently reached a similar level of understanding as the disciples of the Lee-model around 1970, who were fascinated by imaginary bare couplings required to obtain real couplings in the renormalized Hamilton operator of the Lee-model. In \cite{Znojil:2001} one does not only find discussed the concept of {\em pseudo-Hermiticity}, {\em pseudo-unitarity} and {\em indefinite metric} (i.e.\ $P$). It is also explained under the headline {\em Spontaneous Broken {\cal PT} Symmetry} the situation, when the Hamilton operator develops complex ghosts, i.e.\ pairs of mutually complex conjugate complex-valued eigenvalues associated to a biorthogonal eigenbasis of states. Meanwhile there appeared several important new contributions to the field, the enumeration of which would go far beyond the scope of this work. We just want selectively refer here to contributions e.g.\ to the formalism of {\em pseudo-Hermiticity} (e.g.\ \cite{Mostafazadeh:2003iz,scolarici2004,weigert2004} and references therein), {\em spontaneous breakdown of PT-symmetry} (e.g.\ \cite{Weigert:2003pn,levai2004,Dorey:2001hi}), {\em PT-symmetric potentials} (e.g.\ \cite{Levai:2000di}), {\em PT-symmetric matrix models} \cite{Ahmed:2004nm,Bender:2003gu,Bender:2002vv,Bender:2004ss}, {\em reality proofs for spectra of PT-symmetric Hamilton operators} (e.g.\ \cite{Dorey:2003sq,caliceti2004a}), {\em the C- and CPT-operator in PT-symmetric QM} (e.g.\ \cite{Bender:2002vv,Ahmed:2004nm,Bender:2004ss,Bender:2004by,caliceti2004b}), {\em PT-symmetric supersymmetry} (e.g.\ \cite{caliceti2004b,Znojil:2000nh,Znojil:2000fr,Dorey:2001hi}), and {\em generalized creation/annihilation operators} (e.g.\ \cite{Znojil:2000nh,Bagchi:2001gj}).}. 
Due to its intimate relation to the Lee-model the researchers investigating PT-symmetric QT will have to answer the same questions as were raised in the context of the Lee-model, i.e.\ the questions about unitarity, analyticity, Poincar\'{e}-covariance, causality, locality in time-dependent PT-symmetric scattering problems. Like in the foregoing discussion the solution to the questions is provided in terms of (A)CQT, the spacial representation of which is shown below to correspond to the {\em holomorphic representation} in complex analysis, while PT-symmetric QT appears just a subset of (non-Hermitian) (A)CQT. 

\section{(Anti)Causal Quantum Field Theory}
\subsection{(Anti)Causal Klein-Gordon Theory: the Nakanishi Model of 1972}
As mentioned above, N.\ Nakanishi \cite{Kleefeld:Nakanishi:wx,Kleefeld:Nakanishi:1972pt} investigated in 1972 a Lagrangean (See also \cite{froi1959,Lee:iw}!) for a KG field $\phi (x)$ with complex mass $M := m - \frac{i}{2} \, \Gamma$ (and the Hermitian conjugate field $\phi^+(x)$)\footnote{The formalism$\;\!$was$\;\!$1999-2000 independently rederived by the author (see e.g.\ \cite{Kleefeld:2002au,Kleefeld:2002gw,Kleefeld:2001xd}).}. Here we want to introduce immediately isospin and to consider a set of $N$ equal complex mass KG fields $\phi_r (x)$ ($r = 1,\ldots,N$) (i.e.\ a charged ``Nakanishi field'' with isospin $\frac{N-1}{2}$) described by the Larangean
\begin{equation}
{\cal L} (x) = \!\sum\limits_{r} \!\left\{
\frac{1}{2} \left( (\partial \phi_r (x) )^2  - M^2 \, \phi_r (x)^2 \right)\! + 
\frac{1}{2} \left( (\partial \phi_r^+ (x) )^2  - M^{\ast \, 2} \, \phi_r^+ (x)^2 \right) \right\}\,.
\end{equation}
The Lagrange equations of motion for the causal and anticausal ``Nakanishi field'' $\phi_r (x)$ and $\phi_r^+ (x)$, i.e.\ $(\,\partial^2 + M^2) \,\phi_r (x) = 0$ and \mbox{$(\,\partial^2 + M^{\ast \, 2}) \,\phi_r^+ (x) = 0$},  are solved by a Laplace-transform. The result is:
\begin{eqnarray} \phi_r (x) & = & \int 
\frac{d^3 p}{(2\pi )^3 \; 2\, \omega \, (\vec{p}\,)}
\Big[ \, 
 a \, (\vec{p} , r ) \, e^{\displaystyle - \, i p x}  +
 c^+ (\vec{p} , r ) \, e^{\displaystyle i p x}
\,\Big] \Big|_{p^0 = \omega(\vec{p}\,)} \; , \nonumber \\
\phi_r^+ (x) & = & \int
\frac{d^3 p}{(2\pi )^3 2\omega^\ast (\vec{p}\,)}
\Big[ 
  c (\vec{p} , r ) \,e^{\displaystyle - i p^\ast x} +
 a^+ (\vec{p} , r ) \,e^{\displaystyle i p^\ast x}
\Big] \Big|_{p^0 = \omega(\vec{p})}\; , 
\end{eqnarray}
where we defined $a(\vec{p}\,):=a(p)|_{p^{0} = \omega (\vec{p}\,)}$ and $c^+(\vec{p}\,):=a(-p)|_{p^{0} = \omega (\vec{p}\,)}$ on the basis of the complex ``frequency'' $\omega(\vec{p}):=\sqrt{\vec{p}^{\,2}+M^{\,2}}$ ($\omega(\vec{0}):=M$) \footnote{To obtain this result we had to use a delta-distribution ``$\delta (p^2 - M^2)$'' for complex arguments which has been illuminated by N.\ Nakanishi \cite{Kleefeld:Nakanishi1958,Kleefeld:Nakanishi:1972pt}. Nowadays it may be embedded in the framework of (tempered) Ultradistributions \cite{Kleefeld:Bollini:1998en}.}. The theory is quantized by claiming {\em Canonical equal-real-time commutation relations} \footnote{The standard Canonical conjugate momenta to the (anti)causal fields $\phi_r (x)$ and $\phi_r^+ (x)$ are given by $\Pi_r (x) := \delta \,{\cal L}_{0} (x)/\delta (\partial_0 \,\phi_r (x)) = \partial_0 \,\phi_r (x)$ and $\Pi_r^+ (x) := \delta \,{\cal L}_{0} (x)/\delta (\partial_0 \,\phi_r^+ (x)) = \partial_0 \,\phi_r^+ (x)$.}. The non-vanishing commutation relations in configuration space are (with $r,s = 1,\ldots,N$) $[ \, \phi_r (\vec{x},t) , \Pi_s (\vec{y},t) \, ] =  i\, \delta^{\, 3} (\vec{x} - \vec{y}\,) \; \delta_{rs}$ and
 $[ \, \phi^+_r (\vec{x},t) , \Pi^+_s (\vec{y},t) \, ]  =  i\, \delta^{\, 3} (\vec{x} - \vec{y}\,) \; \delta_{rs}$.
The resulting non-vanishing momentum-space commutation relations, which display an {\em indefinite metric}\footnote{An indefinite metric should not surprise, as the space-time metric is $(+,-,-,-)$!}, are ($r,s = 1,\ldots,N$):
\begin{eqnarray} {[ \, a \, (\vec{p},r) \; , \; c^+ (\vec{p}^{\,\,\prime},s) \, ]} & = & (2\pi)^3 \, 2 \; \omega \,(\vec{p}\,)\;\, \delta^{\, 3} (\vec{p} - \vec{p}^{\,\,\prime}\,) \; \delta_{rs} \; ,\nonumber \\
{ [ \, c \, (\vec{p},r) \; , \; a^+ (\vec{p}^{\,\,\prime},s) \, ] } & = & (2\pi)^3 \, 2 \; \omega^\ast(\vec{p}\,)\; \delta^{\, 3} (\vec{p} - \vec{p}^{\,\,\prime}\,) \; \delta_{rs} \; .
\end{eqnarray}
The Hamilton operator is derived by a standard Legendre transform \cite{Kleefeld:2003zj,Kleefeld:2001xd}:
\begin{eqnarray} H & = &
\sum\limits_{r} \;\int\! \!
\frac{d^3 p}{(2\pi )^3 \; 2\, \omega \, (\vec{p}\,)}
\;\;
\frac{1}{2}\;\, \omega \,(\vec{p}\,)\;\, \Big( 
c^+ (\vec{p},r) \; a (\vec{p},r) +
a\, (\vec{p},r) \; c^+ (\vec{p},r) \; \Big) \nonumber \\
 & + & \sum\limits_{r} \,\;\int\! \!
\frac{d^3 p}{(2\pi )^3 \; 2\, \omega^\ast (\vec{p}\,)}
\;
\frac{1}{2}\; \omega^\ast (\vec{p}\,)\; \Big( 
a^+ (\vec{p},r) \; c\, (\vec{p},r) +
c \, (\vec{p},r) \; a^+ (\vec{p},r) \; \Big) \, . \quad \label{nakham1}
\end{eqnarray}
The ``Nakanishi-KG propagator'' $i \, \Delta_N (x-y)$ is obtained by {\em real-time ordering} of causal KG fields \cite{Kleefeld:Nakanishi:wx,Kleefeld:2001xd,Kleefeld:Bollini:1998hj}, i.e.\ $i\, \Delta_N (x-y) \;  \delta_{rs} \equiv \left<\!\left<0\right|\right.T\,[\,\phi_r\,(x)\, \phi_s\, (y)\,]\,\left|0\right>$ \footnote{Explicitly we obtain: $i\,\Delta_N (x-y) \stackrel{!}{=} \int\!\frac{d^{\,4}p}{(2\,\pi)^4}\; e^{-\,i\,p (x-y)} \,\frac{i}{p^2 - M^2}$.
The anticausal ``Nakanishi-KG propagator'' is obtained by Hermitian conjugation or by a vacuum expectation value of an {\em anti-real-time ordered} product of two anticausal fields.
For intermediate states with complex mass these propagators lead to {\em Poincar\'{e} covariant} results, {\em if} causal and anticausal states do not interact. At each interaction vertex coupling to {\em intermediate} complex mass fields there holds {\em exact} 4-momentum conservation. Only if complex mass fields with finite $\Gamma$ appeared as asymptotic states or causal and anticausal states could interact, then Poincar\'{e} covariance would be violated!}. 

\section{Lorentz Transformations in (Anti)Causal Quantum Theory}
A Lorentz transformation $\Lambda^{\mu}_{\,\;\nu}$ for a given (symmetric) metric $g_{\mu\nu}$ is defined by $\Lambda^{\mu}_{\,\;\rho} \; g_{\mu\nu} \; \Lambda^{\nu}_{\,\;\sigma} = g_{\rho\sigma}$. Let $n^{\mu}$ be a timelike unit 4-vector $(n^2 = 1)$ and $\xi^{\mu}$ an {\em arbitrary complex} 4-vector with $\xi^2 \not= 0$. We want to construct \cite{Kleefeld:2002au,Kleefeld:2002gw,Kleefeld:2001xd} a Lorentz transformation $\Lambda^{\mu}_{\,\;\nu}(\xi/\sqrt{\xi^2}\,,n)^{\uparrow}$ relating the 4-vector $\xi^{\mu}$ with its ``restframe'', i.e. $\xi^{\, \mu} = \Lambda^{\,\mu}_{\,\;\,\nu} (\xi/\sqrt{\xi^2}\,,n)^{\uparrow} \; n^{\,\nu} \sqrt{\xi^2}$ and $n_{\,\nu} \sqrt{\xi^2} = \xi_{\, \mu} \, \Lambda^{\,\mu}_{\,\;\,\,\nu} (\xi/\sqrt{\xi^2}\,,n)^{\uparrow}$. The independent quantity $\Lambda^{\mu}_{\,\;\nu}(\xi/\sqrt{\xi^2}\,,n)^{\downarrow}\equiv - \Lambda^{\mu}_{\,\;\nu}(- \xi/\sqrt{\xi^2}\,,n)^{\uparrow}$ being also a Lorentz transformation is simultaneously admissible, as $\xi^{\, \mu} = \Lambda^{\,\mu}_{\,\;\,\nu} (\xi/\sqrt{\xi^2}\,,n)^{\downarrow} \; n^{\,\nu} \sqrt{\xi^2}$ and $n_{\,\nu} \sqrt{\xi^2} = \xi_{\, \mu} \, \Lambda^{\,\mu}_{\,\;\,\,\nu} (\xi/\sqrt{\xi^2}\,,n)^{\downarrow}$. In a first step we define now the {\em inversion matrix} $P^{\,\mu}_{\,\;\,\nu} := 2\, n^{\,\mu} \, n_{\,\nu} - \, g^{\,\mu}_{\,\;\,\nu}$. Then we note: if there is a solution $\Lambda^{\mu}_{\,\;\nu}(\xi/\sqrt{\xi^2}\,,n)^{\uparrow}$ ($\Lambda^{\mu}_{\,\;\nu}(\xi/\sqrt{\xi^2}\,,n)^{\downarrow}$) of the defining equation $\Lambda^{\mu}_{\,\;\rho} \; g_{\mu\nu} \; \Lambda^{\nu}_{\,\;\sigma} = g_{\rho\sigma}$ relating the 4-vector $\xi^{\mu}$ with its ``restframe'', then $\Lambda^{\mu}_{\,\;\beta}(\xi/\sqrt{\xi^2}\,,n)^{\uparrow} P^{\,\beta}_{\,\;\,\nu}$ ($\Lambda^{\mu}_{\,\;\beta}(\xi/\sqrt{\xi^2}\,,n)^{\downarrow} P^{\,\beta}_{\,\;\,\nu}$) will be also a such a solution, respectively. Keeping these remaining solutions in mind we find --- after employing some general ansatz for $\Lambda^{\mu}_{\,\;\nu}(\xi/\sqrt{\xi^2}\,,n)$ --- {\em four} independent solutions $\Lambda^{\mu}_{\,\;\nu}(\xi/\sqrt{\xi^2}\,,n)^{\,\uparrow}_{\pm}$ and $\Lambda^{\mu}_{\,\;\nu}(\xi/\sqrt{\xi^2}\,,n)^{\,\downarrow}_{\pm}$ being related by $\Lambda^{\mu}_{\,\;\nu}(\xi/\sqrt{\xi^2}\,,n)^{\,\uparrow}_{\pm} \equiv - \Lambda^{\mu}_{\,\;\nu}(- \xi/\sqrt{\xi^2}\,,n)^{\,\downarrow}_{\pm}$, which solve the defining equation $\Lambda^{\mu}_{\,\;\rho} \; g_{\mu\nu} \; \Lambda^{\nu}_{\,\;\sigma} = g_{\rho\sigma}$ and relate the 4-vector $\xi^{\mu}$ with its ``restframe''\footnote{Explicitly we find:
$\Lambda^{\mu}_{\,\;\nu}(\xi/\sqrt{\xi^2}\,,n)^{\,\uparrow}_{\pm} = \mp \, \left\{ g^{\,\mu}_{\,\;\,\rho} - \, \frac{\sqrt{\xi^2}}{\sqrt{\xi^2} \pm \xi \cdot n} \, \Big[ n^{\,\mu} \pm \frac{\xi^{\,\mu}}{\sqrt{\xi^2}} \Big] \Big[ n_{\,\rho} \pm \frac{\xi_{\,\rho}}{\sqrt{\xi^2}} \Big] \, 
\right\}  P^{\,\rho}_{\,\;\,\nu}$.}.
$\Lambda^{\mu}_{\,\;\nu}(\xi/\sqrt{\xi^2}\,,n)^{\,\uparrow}_{+}$, $\Lambda^{\mu}_{\,\;\nu}(\xi/\sqrt{\xi^2}\,,n)^{\,\uparrow}_{-}$, $\Lambda^{\mu}_{\,\;\nu}(\xi/\sqrt{\xi^2}\,,n)^{\,\downarrow}_{+}$ and $\Lambda^{\mu}_{\,\;\nu}(\xi/\sqrt{\xi^2}\,,n)^{\,\downarrow}_{-}$ represent the well known orthochronous-proper, orthochronous-improper, nonorthochronous-proper, and nonorthochronous-improper Lorentz transformations, respectively \footnote{Note that $\Lambda^{\mu}_{\,\;\nu}(n,n)^{\,\uparrow}_{+} = - \Lambda^{\mu}_{\,\;\nu}(- n,n)^{\,\downarrow}_{+} = g^{\mu}_{\,\;\nu}$, $\Lambda^{\mu}_{\,\;\nu}(n,n)^{\,\uparrow}_{-} =- \Lambda^{\mu}_{\,\;\nu}(- n,n)^{\,\downarrow}_{-} = P^{\,\mu}_{\,\,\;\nu}$, and $P^{\,\mu}_{\,\;\,\alpha} \, \Lambda^{\alpha}_{\,\;\beta}(n^\prime,n)^{\,\uparrow}_{\pm} \, P^{\,\beta}_{\,\;\,\nu} = \Lambda^{\mu}_{\,\;\nu}(P \, n^\prime,n)^{\,\uparrow}_{\pm}$, $P^{\,\mu}_{\,\;\,\alpha} \, \Lambda^{\alpha}_{\,\;\beta}(n^\prime,n)^{\,\downarrow}_{\pm} \, P^{\,\beta}_{\,\;\,\nu} = \Lambda^{\mu}_{\,\;\nu}(P \, n^\prime,n)^{\,\downarrow}_{\pm}$ for $n^\prime = \xi/\sqrt{\xi^2}$.}! Orthochronous-proper Lorentz transformations respect $\Lambda^{\mu}_{\,\;\alpha}(n^{\prime\prime},n^\prime)^{\,\uparrow}_{+} \Lambda^{\alpha}_{\,\;\nu}(n^\prime,n)^{\,\uparrow}_{+}=\Lambda^{\mu}_{\,\;\nu}(n^{\prime\prime},n)^{\,\uparrow}_{+}$ for $n^2=n^{\prime 2}=n^{\prime\prime 2}=1$. In the following text we use the abbreviation $\Lambda^{\mu}_{\,\;\nu}(\xi)\equiv\Lambda^{\mu}_{\,\;\nu}(\xi/\sqrt{\xi^2}\,,n)^{\,\uparrow}_{+}$. To obtain Lorentz-boost matrices for a particle with {\em complex} mass $M$ and 3-momentum $\vec{p}\,$ being on its {\em complex mass-shell} $p^2 = M^2$, we simply have to set $\xi^\mu=(\omega(\vec{p}\,),\vec{p}\,)^\mu$ and to remember, that we live in a world with a metric $+,-,-,-$ \footnote{As argued in the context of the Nakanishi-KG propagator (A)CQT is Poincar\'{e} invariant even for intermediate complex-mass states with finite $\Gamma$, if the imaginary part of the complex masses of particles appearing as asymptotic states are infinitesimally close to zero and causal and anticausal states do not interact. In (A)CQT {\em causal} states are Lorentz transformed by $\Lambda^{\mu}_{\,\;\nu}(p/(p^2)^{1/2}\,,n)^{\,\uparrow}_{+}$, while {\em anticausal} states are Lorentz transformed by $(\Lambda^{\mu}_{\,\;\nu}(p/(p^2)^{1/2}\,,n)^{\,\uparrow}_{+})^\ast$. Hence, in (A)CQT states are --- against our original expectations \cite{Kleefeld:2003dx,Kleefeld:2003zj} ---- {\em not} representations of the covering group of the {\em complex} Lorentz group $L_+(C)$ (or, more generally, the covering group of the respective Poincar\'{e} group) \cite{Greenberg:2003nv,Streater:1964} which links continuously $\Lambda^{\mu}_{\,\;\nu}(p/(p^2)^{1/2}\,,n)^{\,\uparrow}_{+}$ (containing the unitity $g^{\mu}_{\,\;\nu}$) and $\Lambda^{\mu}_{\,\;\nu}(p/(p^2)^{1/2}\,,n)^{\,\downarrow}_{+}$ (containing the PT-transformation $-\,g^{\mu}_{\,\;\nu}$).}.
\section{The (Anti)Causal Dirac Theory: the Lee-Wick Model of 1970}
The causal Dirac equation and its relatives obtained by Hermitian conjugation/transposition are \cite{Kleefeld:2002au,Kleefeld:2002gw,Kleefeld:1998yj,Kleefeld:1998dg,Kleefeld:thesis1999} $( i \! \stackrel{\;\,\rightarrow}{\partial\!\!\!/} - \, M ) \; \psi_r (x) = 0$, $( i \! \stackrel{\;\,\rightarrow}{\partial\!\!\!/} - \, \bar{M} ) \; \psi_r^c (x) = 0$, $\overline{\psi_r^c} (x) \; ( - \, i \! \stackrel{\;\,\leftarrow}{\partial\!\!\!/} - \, M ) = 0$, and $\bar{\psi}_r (x) \; ( - \, i \! \stackrel{\;\,\leftarrow}{\partial\!\!\!/} - \, \bar{M} ) = 0$ (with $M := m - \frac{i}{2} \, \Gamma$ and $\bar{M}:= \gamma_0 \, M^+ \gamma_0$). Note that $\;r=1, \ldots, N$ is an isospin index and $\psi_r(x)$, $\bar{\psi}_r (x)$, $\psi_r^c (x)= C \, \gamma_0 \,\psi_r^\ast (x)$, $\overline{\psi_r^c} (x)= \psi_r^T(x) \, C$ are Grassmann fields.
The underlying Lagrangean is given by \cite{Kleefeld:2002au,Kleefeld:2002gw,Kleefeld:1998yj,Kleefeld:1998dg,Kleefeld:thesis1999} (see also \cite{Lee:iw}) ($N=1$ yields neutrinos!)\footnote{In (A)CQT antiparticles are isospin partners of particles. Charged pions ($\pi^+$, $\pi^-$) are e.g.\ represented by the isospin combination $\pi_\pm(x) = (\pi_1(x) \pm i \, \pi_2(x))/\sqrt{2}$, while the positron ($e^+$) and electron ($e^-$) are represented by $e_\pm(x) = (e_1(x) \pm i \, e_2(x))/\sqrt{2}$. $\pi_1(x)$, $\pi_2(x)$ or $e_1(x)$, $e_2(x)$ are non-Hermitian fields describing, respectively, pairs of causal neutral particles with equal complex mass. This holds for Bosons and Fermions. It leads --- like in the Bosonic case --- to the fact that the anti-Fermions have the same intrinsic parity as the Fermions, contrary to the conclusions of V.B.\ Berestetskii obtained in HQT and mentioned in the introduction. In spite of this feature (A)CQT reproduces exactly the high precision results of QED.}
\begin{equation} {\cal L} (x) = \sum\limits_r \,\frac{1}{2} \left( \; \overline{\psi_r^c} (x) \, ( \frac{1}{2} \, i \! \stackrel{\;\,\leftrightarrow}{\partial\!\!\!/} \! -  M ) \, \psi_r (x) \; + \; \bar{\psi}_r (x) \, ( \frac{1}{2} \, i \! \stackrel{\;\,\leftrightarrow}{\partial\!\!\!/} \! -  \bar{M} ) \, \psi_r^c (x) \; \right) \; .
\end{equation}
4-spinors $u(p,s)\equiv v(-p,s)$ in complex 4-momentum space ($s=\pm \, \frac{1}{2}$) are introduced by the defining equation $( \, p\!\!\!/ -  \sqrt{p^2} \; ) \; u(p,s) \;  = \; 0\; \Leftrightarrow \;  \overline{u^c}(p,s) \, ( \, - \, p\!\!\!/ -  \sqrt{p^2} \; ) \; = \; 0$.
The spinors are normalized according to $\mbox{sgn[Re}(p^0) \,] \; \sum_s \, u(p,s) \; \overline{v^c}(p, s) =  \; p\!\!\!/ +  \sqrt{p^2}$ for $\mbox{Re}[p^0]\;\ne\;0$. Equations of motions are solved by a Laplace-transformation \footnote{With $b_r(-p,s)\equiv d_r^+(p,s)$, $b_r(\vec{p},s)\equiv b_r(p,s)|_{p^0 = \omega (\vec{p}\,)}$, $d^+_r(\vec{p},s)\equiv d^+_r(p,s)|_{p^0 = \omega (\vec{p}\,)}$ we obtain:
\begin{eqnarray} \psi_r (x) & = & \sum\limits_s \int \frac{d^3p}{(2\pi)^3 \; 2 \, \omega (\vec{p}\,)} \;\; [ \,e^{-\, i \,  p\cdot x} \;\,\, b_{\,r} (p, s) \, u(p,s) \; +  e^{+ \, i \,  p\cdot x} d^+_{\,r} (p, s) \, v(p,s)\,] \big|_{p^0 = \omega (\vec{p}\,)} \, , \nonumber \\
 \psi_r^c (x) & = & \sum\limits_s \int \frac{d^3p}{(2\pi)^3 \; 2 \, \omega^\ast (\vec{p}\,)} \, [ \,e^{+\, i \,  p^\ast\cdot x} b^+_{\,r} (p, s) \, u^c(p,s) + e^{- \, i \,  p^\ast\cdot x} d_{\,r} (p, s) \, v^c(p,s)\,] \big|_{p^0 = \omega (\vec{p}\,)} \, , \nonumber \\
 \overline{\psi_r} (x) & = & \sum\limits_s \int \frac{d^3p}{(2\pi)^3 \; 2 \, \omega^\ast (\vec{p}\,)} \, [ \,e^{+ i \,  p^\ast \cdot x} \, b^+_{\,r} (p, s) \, \bar{u}(p,s) \; + e^{- \, i \,  p^\ast\cdot x} d_{\,r} (p, s) \, \bar{v}(p,s)\,] \big|_{p^0 = \omega (\vec{p}\,)} \, , \nonumber \\
 \overline{\psi^c_r} (x) & = & \sum\limits_s \int \frac{d^3p}{(2\pi)^3 \; 2 \, \omega (\vec{p}\,)} \, \, \;[ \,e^{- i \,  p \cdot x} \;\,\,  b_{\,r} (p, s) \, \overline{u^c}(p,s) + e^{+ \, i \,  p \cdot x} d^+_{\,r} (p, s) \, \overline{v^c}(p,s)\,] \big|_{p^0 = \omega (\vec{p}\,)}\, , \nonumber 
\end{eqnarray}
and $\{b_r(\vec{p},s),d_{r^\prime}^+(\vec{p}^{\,\prime},s^{\,\prime})\}/(2\omega (\vec{p}))= \{d_r(\vec{p},s),b_{r^\prime}^+(\vec{p}^{\,\prime},s^{\,\prime})\}/(2\omega^\ast (\vec{p}))= (2\pi)^3 \delta^3(\vec{p}-\vec{p}^{\,\prime}) \delta_{s s^\prime}\delta_{r r^\prime}, \ldots$}.
Note that the spinors obey the analyticity property $u^c(p,s) \,  = \, u(-\,p^\ast ,s) \,  = \, v(p^\ast ,s)$. 
The (anti)causal Dirac equation is Lorentz covariant due to standard transformation properties of spinors and $\gamma$-matrices, i.e.\ $u(p) = S(\Lambda(p)) \; u(\sqrt{p^2} \; n)$ and $S^{-1} (\Lambda (p)) \, \gamma^{\,\mu} \, S (\Lambda (p)) \, = \, \Lambda^{\,\mu}_{\,\,\;\nu} (p) \; \gamma^{\,\nu}$.
The causal ``Nakanishi-Dirac propagator'' $i \, S_{N} (x-y)_{\alpha\beta}$ of a spin 1/2 Fermion is obtained by {\em standard Fermionic real-time ordering} of causal Dirac fields, i.e.\ $i \, S_{N} (x-y)_{\alpha\beta} \delta_{rs} \equiv \left<\!\left<0\right|\right. T\,[\,(\psi_r (x))_\alpha (\,\overline{\psi^c_s} (y))_\beta\,] \left|0\right>$~\footnote{Explicitly we obtain: $i \, S_{N} (x-y)_{\alpha\beta} \stackrel{!}{=} \int\!\frac{d^{\,4}p}{(2\,\pi)^4}\;e^{-i\,p (x-y)}\; \frac{i}{p^2 - M^2} \, (\not\!p + M)_{\alpha \beta}$.
The anticausal ``Nakanishi-Dirac propagator'' is obtained by Hermitian conjugation or by a vacuum expectation value of a {\em anti-real-time ordered} product of two anticausal Dirac fields.}.
\section{(Anti)Causal Massive and ``Massless'' Vector Fields}
With \footnote{The problem of constructing (anti)causal vector fields is twofold. First one has to be aware that even a ``massless'' (anti)causal field has to be treated as if it were massive due to the at least infinitesimal imaginary part of its complex mass. That even non-Abelian massive vector fields can be treated consistently within QFT without relying on a Higgs mechanism has been shown by Jun-Chen Su in a renormalizable and unitary formalism \cite{Su:1998wy}. Secondly, one has to be able to construct polarization vectors based on a boost of complex mass fields.}
$\vec{e}^{\,\,(i)}\; (i=x,y,z)$ and $\vec{e}^{\,\,(i)} \cdot \vec{e}^{\,\,(j)} = \delta^{ij}$ we define the polarization vectors
$\varepsilon^{\,\mu \, (i)} (p) := \Lambda^\mu_{\;\nu} (p) \; \varepsilon^{\,\nu \, (i)} (\sqrt{p^2},\vec{0}) = \Lambda^\mu_{\;\nu} (p) \, (0,\vec{e}^{\,\,(i)})^{\,\nu}$. In the chosen unitary gauge they obey $p^\mu  \, \varepsilon_\mu^{\,(i)} (p) = 0$, $\varepsilon^{\,\mu \, (i)} (p) \, \varepsilon_\mu^{\, (j)} (p) = - \,\delta^{ij}$, and $\sum_i \varepsilon^{\,\mu \, (i)} (p) \, \varepsilon^{\,\nu \, (i)} (p) =  - g^{\,\mu\nu} + \frac{p^{\mu}\, p^{\nu}}{p^2}$.
Based on these polarization vectors the Bosonic field operators for causal vector fields $V_r^\mu (x)$ ($r=1,\ldots,N=$ isospin index) and anticausal vector fields $(V_r^\mu (x))^+$ are --- as the Nakanishi-KG field --- introduced by $(V_r^\mu (x))^+ = V_r^\mu (x) = \sum\limits_{j} \int \frac{d^3p}{(2\pi)^3 \; 2\, \omega (\vec{p}\,)} \;\;\varepsilon^{\,\mu \, (j)} (p) \; \, [ \; e^{-\, i \,  p\cdot x} \; a_r (p,j) + e^{+ \, i \,  p\cdot x} \; a_r (- \, p,j)] \Big|_{p^{0} = \omega (\vec{p}\,)\,}$. 
\section{(Anti)Causal Quantum Mechanics}
\subsection{``Deducing'' (Anti)Causal Schr\"odinger from (Anti)Causal KG Theory} \label{maxb1}
The causal KG equation \mbox{$(\partial^2 + M^2 )\, \phi (x)\, = 0$} is second order in time, which gets transparent by factorizing the respective {\em causal KG differential operator} according to $(\partial^2 + M^2 ) = (i\,\partial_t - \omega(-\,i\stackrel{\,\rightarrow}{\nabla}))(i\,\partial_t + \omega(-\,i\stackrel{\,\rightarrow}{\nabla}))$ \footnote{The same holds for the adjoint causal KG equation $\phi(x)(\,\stackrel{\leftarrow}{\partial}\!{}^2 + M^2 )=0$ and the respective {\em adjoint causal KG differential operator} $(\,\stackrel{\leftarrow}{\partial}\!{}^2 + M^2 ) = (- i\stackrel{\leftarrow}{\partial}_t - \, \omega(i \stackrel{\leftarrow}{\nabla}))(- i\stackrel{\leftarrow}{\partial}_t + \, \omega(i\stackrel{\leftarrow}{\nabla}))$ being obtained by partial integration and providing us with the left eigensolutions of the causal KG differential operator.
The anticausal KG equation $(\partial^2 + M^{\ast \, 2} )\, \phi^+ (x)\, = 0$ and the adjoint anticausal KG equation $\phi^+(x)(\,\stackrel{\leftarrow}{\partial}\!{}^2 + M^{\ast\,2} )=0$ are obtained by Hermitian conjugation from the adjoint causal KG equation and the causal KG equation, respectively.}. Inspection of the ``factorized'' causal and anticausal KG equations and their adjoints \footnote{$0=\big(i\,\partial_t - \omega(-i\stackrel{\,\rightarrow}{\nabla})\big)\big(i\,\partial_t + \omega(-i\stackrel{\,\rightarrow}{\nabla})\big) \phi(x) = \big(i\,\partial_t - \omega^\ast(i\stackrel{\,\rightarrow}{\nabla})\big)\big(i\,\partial_t + \omega^\ast (i\stackrel{\,\rightarrow}{\nabla})\big) \phi^+(x) = \phi(x) \big(- i\stackrel{\leftarrow}{\partial}_t - \, \omega(i \stackrel{\leftarrow}{\nabla})\big)\big(- i\stackrel{\leftarrow}{\partial}_t + \, \omega(i\stackrel{\leftarrow}{\nabla})\big) = \phi^+(x) \big(- \,i\stackrel{\leftarrow}{\partial}_t - \, \omega^\ast(-\,i\stackrel{\leftarrow}{\nabla})\big)\big(-\,i\stackrel{\leftarrow}{\partial}_t + \, \omega^\ast (-\,i\stackrel{\leftarrow}{\nabla})\big)$.}
yields eight underlying differential equations of first order in time:
\begin{eqnarray} \left(i\,\partial_t - \omega(-\,i\stackrel{\,\rightarrow}{\nabla})\right) \,\phi^{(+)}(x) = 0 \; , & & \left(\phi^{(+)}(x)\right)^+ \left(- \,i\stackrel{\leftarrow}{\partial}_t - \, \omega^\ast(-\,i\stackrel{\leftarrow}{\nabla})\right) = 0 \; , \nonumber \\
\left(i\,\partial_t + \omega(-\,i\stackrel{\,\rightarrow}{\nabla})\right) \,\phi^{(-)}(x) = 0 \; , & & \left(\phi^{(-)}(x)\right)^+ \left(- \,i\stackrel{\leftarrow}{\partial}_t + \, \omega^\ast(-\,i\stackrel{\leftarrow}{\nabla})\right) = 0 \; , \nonumber \\
\phi^{(-)}(x) \left(- i\stackrel{\leftarrow}{\partial}_t - \,\omega(i \stackrel{\leftarrow}{\nabla})\right) = 0 \; , & & \left(i\,\partial_t - \omega^\ast(i\stackrel{\,\rightarrow}{\nabla})\right) \left(\phi^{(-)}(x)\right)^+ = 0 \; , \nonumber \\
\phi^{(+)}(x) \left(- i\stackrel{\leftarrow}{\partial}_t + \, \omega(i \stackrel{\leftarrow}{\nabla})\right) = 0 \; , & & \left(i\,\partial_t + \omega^\ast(i\stackrel{\,\rightarrow}{\nabla})\right) \left(\phi^{(+)}(x)\right)^+ = 0 \; .
\end{eqnarray}
For a non-Hermitian Hamilton operator $H$ we can construct corresponding Schr\"odin-ger equations by replacing the spacial parts of KG differential operators by respective Hamilton operators (Replace e.g.\ $\omega(-\,i\stackrel{\,\rightarrow}{\nabla})$ by $H$, $\omega(i \stackrel{\leftarrow}{\nabla})$ by $\stackrel{\leftarrow}{H}\,$, etc.!). Hence, the (Anti)Causal Schr\"odinger Theory is described by the following eight equations:
\begin{eqnarray} \left(i\,\partial_t - H\,\right) \big| \psi (t) \big> = 0 \; , & & \big< \psi (t) \big| \left(-\,i\stackrel{\leftarrow}{\partial}_t - \, H^+\right) = 0 \; , \nonumber \\
\left(i\,\partial_t + H\,\right)\big| \tilde{\psi} (t) \big>  = 0 \; , & &\big< \tilde{\psi} (t) \big| \left(-\,i\stackrel{\leftarrow}{\partial}_t + \, H^+\right) = 0 \; , \nonumber \\
\big<\!\big< \tilde{\psi} (t) \big|\left(- i\stackrel{\leftarrow}{\partial}_t - \stackrel{\,\leftarrow}{H}\,\right) = 0 \; , & & \left(i\,\partial_t - \,\big(\!\stackrel{\,\leftarrow}{H} \big)^+\right) \big| \tilde{\psi} (t) \big>\!\big> = 0 \; , \nonumber \\
\big<\!\big< \psi (t) \big| \left( - i\stackrel{\leftarrow}{\partial}_t + \stackrel{\,\leftarrow}{H}\,\right) = 0 \; , & & \left(i\,\partial_t + \,\big(\!\stackrel{\,\leftarrow}{H} \big)^+\right) \big| \psi (t) \big>\!\big> = 0 \; .\label{schreq1}
\end{eqnarray}
Causal equations (left column) are related to corresponding anticausal equations (right column) by Hermitian conjugation.
From the causal equations (left column) we can derive the continuity equations $\partial_t \big<\!\big< \tilde{\psi} (t) \big| \psi (t) \big> = - \, i \, \big<\!\big< \tilde{\psi} (t) \big|\big(H \,- \stackrel{\,\leftarrow}{H}\! \big)\big| \psi (t) \big>$ and
$\partial_t \big<\!\big< \psi (t) \big| \tilde{\psi} (t) \big> = + \, i \, \big<\!\big< \psi (t) \big|\big(H \,- \stackrel{\,\leftarrow}{H}\!\big)\big| \tilde{\psi} (t) \big>$ \footnote{The respective anticausal continuity equations are obtained by Hermitian conjugation.}. The right hand side of these equations leads for standard --- even non-Hermitian --- Hamilton operators being quadratic in the momentum at most to (spacial) surface terms!
The corresponding non-vanishing --- in general {\em complex} --- densitities $\big<\!\big< \tilde{\psi} (t) \big| \psi (t) \big>$ and  $\big<\!\big< \psi (t) \big| \tilde{\psi} (t) \big>$ being in the standard way related to --- in general {\em complex} --- conserved charges may be interpreted as {\em complex} probability densities \cite{berggren1970} which replace Born's \cite{Pais:1982we} suggested ansatz $\big< \psi (t) \big| \psi (t) \big>$ for a {\em real} probability density. We refer here also to other interesting attempts to change or extend Born's scalar product, e.g.\ \cite{bollini1996,moiseyev1998}!
\subsection{The ``Representation Free'' (Anti)Causal Harmonic Oscillator} \label{repfree1}
The Hamilton operator of the (Anti)Causal Harmonic Oscillator in 1-dim.\ QM is given in analogy to the Hamilton operator Eq.\ (\ref{nakham1}) of the Nakanishi model by:
\begin{equation} H \; = \; H_C + H_A \; = \; \frac{1}{2} \, \omega \, [c^+, a\, ]_\pm + \frac{1}{2} \, \omega^\ast \, [a^+, c\, ]_\pm \qquad (H_A = H^+_C)\end{equation}
($\pm$ for Bosons/Fermions \footnote{The Fermionic case we tend to denote by $H = \frac{1}{2} \omega [d^+, b ] + \frac{1}{2} \omega^\ast [b^+, d ]$ with $\{\,b, d^+ \} = 1$ etc.\,.}) \cite{Kleefeld:Nakanishi:1972pt,Kleefeld:2003zj,Kleefeld:2002gw,Kleefeld:2001xd,Kleefeld:1998yj,Kleefeld:1998dg,Kleefeld:thesis1999,Kleefeld:Kossakowski2002}. 

\noindent The relevant (anti)commuation relations are:
\begin{eqnarray} \left( \begin{array}{cc} {[c,c^+]_\mp} & {[c,a^+]_\mp} \\
{[a,c^+]_\mp} & {[a,a^+]_\mp} \end{array}\right) \quad = \quad \left( \begin{array}{cc} 0 & 1 \\
1 & 0 \end{array}\right) \quad = \quad \mbox{``indefinite metric''} \; .
\end{eqnarray}
Of course there also holds:
\begin{eqnarray} \left( \begin{array}{cc} {[c,c]_\mp} & {[c,a]_\mp} \\
{[a,c]_\mp} & {[a,a]_\mp} \end{array}\right) \quad = \quad \left( \begin{array}{cc} {[c^+,c^+]_\mp} & {[c^+,a^+]_\mp} \\
{[a^+,c^+]_\mp} & {[a^+,a^+]_\mp} \end{array}\right) \quad = \quad \left( \begin{array}{cc} 0 & 0 \\
0 & 0 \end{array}\right) \; .
\end{eqnarray}
Obviously there holds $[H_C,H_A]=0$. One of the important problems is now to find the right and left eigensystem of the Hamilton operator for given energy $E$, i.e.\ to solve the following two (stationary) Schr\"odinger equations $(H-E) \left|\psi\right> = 0$ and $\left<\!\left<\,\psi\right|\right.\! (H-E) = 0$.
The resulting (normalized) {\em normal} right eigenstates $\big|n,m\big>$ and left eigenstates $\big<\!\big<\,n,m\big|$ are given by (Bosons: $n,m\in{\rm I\!N}$${}_0$\,; Fermions: $n,m\in\{0,1\}$)
\begin{equation} \big|n,m\big> = \frac{1}{\sqrt{n!\,m!}} \,\, (c^+)^n (a^+)^m \big|0\big> \; ,\quad
\big<\!\big<n,m\big| = \frac{1}{\sqrt{m!\,n!}} \, \big<\!\big<\,0\big|\, c^m \, a^n \; . \label{hostates1}
\end{equation}
They are solutions of the equations $(H-E_{\,n,m}) \big|n,m\big> = 0$ and $\big<\!\big< n,m\big| (H-E_{\,n,m}) = 0$
for the eigenvalues $E_{n,m} = \omega  (n \pm \frac{1}{2}) + \omega^\ast (m \pm \frac{1}{2})$. The eigenvalues $E_{n,n}$ are obviously {\em real}, while the eigenvalues $E_{m,n}$ and $E_{n,m}$ form a {\em complex conjugate} pair for $n\not=m$ which arises typically for the case of broken ``PT''-symmetry.
The {\em (bi)orthogonal} (here {\em normal}) eigenstates are complete:
\begin{equation} \big<\!\big< n^\prime,m^\prime\big|n,m\big> = \delta_{n^\prime n} \, \delta_{m^\prime m} \; , \quad
\sum\limits_{n,m} \big|n,m\big>\big<\!\big<n,m\big| = \mbox{\bf 1} \; .
\end{equation}
We note that one can construct (e.g.\ \cite{gupta1957,Kleefeld:2003zj,Kleefeld:2003dx,nagy1960b,nagy1966,Mostafazadeh:2004ph,Habara:2003cz}) for the (anti)causal Harmonic Oscillator also an {\em abnormal} basis of eigenstates being complete with respect to negative (complex) energy eigenvalues by applying annihilation (creation) operators on a newly defined {\em dual vacuum} state $\left|\bar{0}\right>$ ($\left<\!\left< \bar{0}\right|\right.$) annihilating creation (annihilation) operators and creating annihilation (creation) operators, respectively. 
\subsection{The (Anti)Causal Harmonic Oscillator in Holomorphic Representation}
As in the traditional formulation of QM it is now natural to look for a spacial representation of the ``representation free'' oscillator introduced above. The particular complication induced by the existence of {\em two} types of annihilation operators~\mbox{($a$, $c$)} and creation operators ($c^+$, $a^+$) indicating a {\em doubling} of the degrees of freedom compared to the traditional Harmonic Oscillator is overcome by replacing the originally {\em real} spacial variable $x$ by a {\em complex} variable $z$ and its complex conjugate $z^\ast$ \footnote{M.\ Znojil \cite{Znojil:2001} achieves a complexification of a real coordinate $x$ by an overall shift $x\rightarrow x-i\delta$.}. This replacement of functions of one real argument $f(x)$ ($x\in{\rm I\!R}$) by respective functions $f(z,z^\ast)$ of complex arguments $z,z^\ast\in{\rm\bf C}$ is well known and used in complex analysis \cite{ablowitz2003} under the terminology ``holomorphic representation'' (e.g.\ \cite{Faddeev:1980be,Kleefeld:Chruscinski2002,moiseyev1998}). Replacing the right and left eigenstates $\big|x\big>$ and $\big<x\big|$ of the position operator by respective states in holomorphic representation $\big|z,z^\ast\big>$ and $\big<\!\big<z,z^\ast\big|$, we can --- analogously to traditional QM --- denote the Schr\"odinger equations and their adjoints given e.g.\ in the first column of Eq.\ (\ref{schreq1}) in their holomorphic representation by:
\begin{eqnarray} + i \,\partial_t \,\big<\!\big<z,z^\ast\big|\psi(t)\big> & = & \int dz^\prime dz^{\prime\ast} \, \big<\!\big<z,z^\ast\big|H\big|z^\prime,z^{\prime\ast}\big>\big<\!\big<z^\prime,z^{\prime\ast}\big|\psi(t)\big> \; , \nonumber \\
- i \,\partial_t \,\big<\!\big<z,z^\ast\big|\tilde{\psi}(t)\big> & = & \int dz^\prime dz^{\prime\ast} \, \big<\!\big<z,z^\ast\big|H\big|z^\prime,z^{\prime\ast}\big>\big<\!\big<z^\prime,z^{\prime\ast}\big|\tilde{\psi}(t)\big> \; , \nonumber \\
- i \,\partial_t \,\big<\!\big<\tilde{\psi}(t)\big|z,z^\ast\big> & = & \int dz^\prime dz^{\prime\ast} \, \big<\!\big<\tilde{\psi}(t)\big|z^\prime,z^{\prime\ast}\big>\big<\!\big<z^\prime,z^{\prime\ast}\big|\!\stackrel{\,\leftarrow}{H}\big|z,z^\ast\big> \; , \nonumber \\
+ i \,\partial_t \,\big<\!\big<\psi(t)\big|z,z^\ast\big> & = & \int dz^\prime dz^{\prime\ast} \, \big<\!\big<\psi(t)\big|z^\prime,z^{\prime\ast}\big>\big<\!\big<z^\prime,z^{\prime\ast}\big|\!\stackrel{\,\leftarrow}{H}\big|z,z^\ast\big> \; .
\end{eqnarray}
The spacial integration contours $\int dz^\prime\, dz^{\prime\ast}$ are to be performed such that there holds the generalized completeness relation $\int dz \, dz^\ast \,\big|z,z^\ast\big>\big<\!\big<z,z^\ast\big| = \mbox{\bf 1}$. 
Inversely, by the used notation it is understood that there holds a generalized orthogonality relation $\big<\!\big<z,z^\ast\big|z^\prime,z^{\prime\ast}\big> = \delta(z-z^\prime)\,\delta(z^\ast - z^{\prime\ast})$, in which the $\delta$-distributions for complex arguments are assumed to exist with respect to the chosen integration contours mentioned above. The holomorphic representation of the (Anti)Causal {\em Bosonic} Harmonic Oscillator with $H = H_C + H_A = \frac{1}{2} \, \omega  \{c^+, a\} + \frac{1}{2} \, \omega^\ast \{a^+, c\}$ is provided by a translational invariant Hamilton operators $\big<\!\big<z,z^\ast\big|H\big|z^\prime,z^{\prime\ast}\big>= H(z,z^\ast) \,\big<\!\big<z,z^\ast\big|z^\prime,z^{\prime\ast}\big> = \big(H_C(z) + H_A(z^\ast)\big) \,\big<\!\big<z,z^\ast\big|z^\prime,z^{\prime\ast}\big>$ and $\big<\!\big<z^\prime,z^{\prime\ast}\big|\!\!\stackrel{\,\leftarrow}{H}\big|z,z^\ast\big> = \big<\!\big<z^\prime,z^{\prime\ast}\big|z,z^\ast\big>\!\stackrel{\,\leftarrow}{H}\!(z,z^\ast) = \big<\!\big<z^\prime,z^{\prime\ast}\big|z,z^\ast\big> \, \big(\!\stackrel{\,\leftarrow}{H}_C\!(z) + \stackrel{\,\leftarrow}{H}_A\!(z^\ast)\big)$ with
\begin{eqnarray} H(z,z^\ast) & = & -\, \frac{1}{2\, M} \,\frac{d^2}{dz^2} + \frac{1}{2} \; M \, \omega^2 \, z^2 -\, \frac{1}{2\, M^\ast} \, \frac{d^2}{dz^{\ast 2}} + \frac{1}{2} \; M^\ast \, \omega^{\ast 2} \, z^{\ast 2} \; , \nonumber \\
\stackrel{\,\leftarrow}{H}\!(z,z^\ast) & = & -\, \frac{1}{2\, M} \stackrel{\leftarrow}{\frac{d^2}{dz^2}} + \frac{1}{2} \; M \, \omega^2 \, z^2 -\, \frac{1}{2\, M^\ast} \stackrel{\leftarrow}{\frac{d^2}{dz^{\ast 2}}} \! + \, \frac{1}{2} \; M^\ast \, \omega^{\ast 2} \, z^{\ast 2} \; .
\end{eqnarray}
Momentum operators $p$ and $p^\ast$ are introduced by $p = - i \, d/dz$ and $p^\ast = - i \, d/dz^\ast$.
Obviously there holds the following correspondence between annihilation/creation operators in the representation free case and in the holomorphic representation: 
\begin{eqnarray} c^+ & \leftrightarrow & \frac{1}{\sqrt{2\,M\,\omega}} \, \left( p + i\, M\,\omega \, z \right) \; , \quad c \; \; \;\leftrightarrow \; \frac{1}{\sqrt{2\,M^\ast\,\omega^\ast}} \, \left( p^\ast -  i\, M^\ast\,\omega^\ast \, z^\ast \right) \; , \nonumber \\
a \; \;& \leftrightarrow & \frac{1}{\sqrt{2\,M\,\omega}} \, \left( p - i\, M\,\omega \, z \right)  \; , \quad a^+ \; \leftrightarrow \; \frac{1}{\sqrt{2\,M^\ast\,\omega^\ast}} \, \left(p^\ast + i\, M^\ast\,\omega^\ast \,z^\ast \right) \; . \quad
\end{eqnarray}
The inverse correspondence is:
\begin{eqnarray} z & \leftrightarrow & \;\;\, \frac{i}{\sqrt{2\,M\, \omega}} \; (a - c^+)\; , \quad 
\,p \;\,=  - \, i \, \frac{d}{dz} \; \;\; \leftrightarrow \; \; \;\;\sqrt{\frac{M\,\omega}{2}} \; (a+c^+) \; , \nonumber \\
 z^\ast & \leftrightarrow & \frac{i}{\sqrt{2\,M^\ast\, \omega^\ast}} \; (c - a^+) \; , \quad p^\ast = - \, i \, \frac{d}{dz^\ast} \; \leftrightarrow \; \sqrt{\frac{M^\ast\,\omega^\ast}{2}} \; (c+a^+) \; . 
\end{eqnarray}
By this correspondence it is straight forward to construct the {\em normal} eigensolutions of the stationary Schr\"odinger equations $H(z,z^\ast )\,\,\big<\!\big< z,z^\ast\big|n,m\big> = E_{n,m} \,\big<\!\big< z,z^\ast\big|n,m\big>$ and $\big<\!\big<n,m\big| z,z^\ast\big> \! \stackrel{\,\leftarrow}{H}\!(z,z^\ast ) = E_{n,m} \,\big<\!\big<n,m\big| z,z^\ast\big>$ of the (Anti)Causal Bosonic Harmonic Oscillator in holomorphic representation. They are given by:
\begin{eqnarray} \lefteqn{\big<\!\big< z,z^\ast\big| n,m\big> 
 \; = \; \frac{1}{\sqrt{n!\,m!}} \,\, \big<\!\big< z,z^\ast\big| (c^+)^n (a^+)^m \big|0\big>} \nonumber \\[1mm]
 & = & i^{\,n+m}\, \sqrt{\frac{|M\omega|}{2^{\,n+m} \, n!\,  m! \, \pi}} \;\, \exp\left(-\,\frac{1}{2}\, ( \xi^2 + \xi^{\ast\,2}\,)\right) \; H_n(\xi )\,H_m(\xi^\ast) \; , \nonumber \\[1mm]
\lefteqn{\big<\!\big< n,m\big|z,z^\ast \big> 
 \; = \; \frac{1}{\sqrt{m!\,n!}} \,\, \big<\!\big< 0\big| \; c^m \; a^n \big|z,z^\ast \big>} \nonumber \\[1mm]
 & = & (-i)^{\,m+n}\, \sqrt{\frac{|M\omega|}{2^{\,m+n} \, m!\,  n! \, \pi}} \;\, \exp\left(-\,\frac{1}{2}\, (\xi^{\ast\,2}+\xi^2\,)\right) \; H_m(\xi^\ast) \, H_n(\xi) \; , 
\end{eqnarray}
with $\xi = z \; \sqrt{M \,\omega}$ and $\xi^\ast = z^\ast \; \sqrt{M^\ast \,\omega^\ast}$. The inverse oscillator length $\sqrt{M \,\omega}$ is here {\em complex} valued \footnote{The Hermite polynomials $H_n(\xi)$ and $H_m(\xi^\ast)$ are defined by: \begin{equation} H_n(\xi) = e^{\,\xi^2\!/2} \left( \xi -\,\frac{d}{d\xi} \right)^n e^{-\,\xi^2\!/2} \; , \; H_m(\xi^\ast) =  e^{\,\xi^{\ast \,2}\!/2} \left( \xi^\ast -\,\frac{d}{d\xi^\ast} \right)^m e^{-\,\xi^{\ast \,2}\!/2} \; . \end{equation}} \footnote{Note that (anti)causal orthonormality relations $\int d\xi \, \exp(-\xi^2) \,H_n (\xi) \,H_{m} (\xi) = 2^n n! \sqrt{\pi}\,\delta_{n m}$ and $\int d\xi^\ast \, \exp(-\xi^{\ast 2}) \,H_n (\xi^\ast) \,H_{m} (\xi^\ast) = 2^n n! \sqrt{\pi}\,\delta_{n m}$
are quite different from the ones of Glauber coherent states \cite{Glauber:1963tx,Faddeev:1980be}: $\int d\xi \,d\xi^\ast \exp(-|\xi|^2) \,\psi^\ast_n (\xi) \, \psi_m(\xi^\ast)/(2\pi i) = \delta_{n m}$ with $\psi_n(\xi)=\xi^n/\sqrt{n!}$.}.

\subsection{Space-Reflection and Time-Reversal in (Anti)Causal Quantum Theory}
Based on the previous considerations it is now straight forward to introduce operations $P$, $T$, ${\cal T}$, $PT$, and $P{\cal T}$ involving reflections of space and/or time, whose effect on $z$, $p$, $z^\ast$, $p^\ast$, $a$, $c$, $a^+$, $c^+$, and the imaginary unit $i=\sqrt{-1}$ is specified in Table \ref{tabop1} \footnote{According to Table \ref{tabop1} the standard space reflection $P$ transforms the holomorphic coordinates $z$ and $z^\ast$ and the respective momenta $p$ and $p^\ast$ in the following way: $z\stackrel{P}{\rightarrow} -\,z$, $p\stackrel{P}{\rightarrow} -\,p$, $z^\ast\stackrel{P}{\rightarrow} -\,z^\ast$, $p^\ast\stackrel{P}{\rightarrow} -\,p^\ast$. These operations result in a transformation of {\em Bosonic} annihilation and creation operators according to the rules $a\stackrel{P}{\rightarrow} -\,a$, $c\stackrel{P}{\rightarrow} -\,c$, $a^+\stackrel{P}{\rightarrow} -\,a^+$, $c^+\stackrel{P}{\rightarrow} -\,c^+$, while the imaginary unit $i=\sqrt{-1}$ remains invariant ($+i\stackrel{P}{\rightarrow} +i$\,).}.
%
%
\begin{table}[t]
\caption{Effect of the operations $P$, $T$, ${\cal T}$, $PT$, and $P{\cal T}$ within {\em Bosonic} systems.} \label{tabop1}
\vspace{2mm}
\small
\begin{center}
\begin{tabular}{|l||l|l|l|l|l|}
\hline
 & $\;\;P$ & $T$ & ${\cal T}$ & $PT$ & $P{\cal T}$ \\
\hline\hline
 $z$      & $- z$      & $+ z$      & $+ z^\ast$ & $-z$      & $-z^\ast$ \\[0.5mm]
 $p$      & $- p$      & $- p$      & $+ p^\ast$ & $+p$      & $-p^\ast$ \\[0.5mm]
 $z^\ast$ & $- z^\ast$ & $+ z^\ast$ & $+ z$      & $-z^\ast$ & $-z$ \\[0.5mm]
 $p^\ast$ & $- p^\ast$ & $- p^\ast$ & $+ p$      & $+p^\ast$ & $-p$ \\[0.5mm]
\hline
 $a$ &   $- a$   & $- c^+$ & $+ a^+$ & $+c^+$ & $-a^+$ \\[0.5mm]
 $c$ &   $- c$   & $- a^+$ & $+ c^+$ & $+a^+$ & $-c^+$ \\[0.5mm]
 $a^+$ & $- a^+$ & $- c$   & $+ a$   & $+c$   & $-a$ \\[0.5mm]
 $c^+$ & $- c^+$ & $- a$   & $+ c$   & $+a$   & $-c$ \\[0.5mm]
\hline
 $+i$ & $+ i$ & $+ i$ & $-i$ & $+ i$ & $-i$ \\[0.5mm]
\hline
\end{tabular}
\vspace{-1mm}
\end{center}
\end{table}
The time-reversal operation is not unique within (Anti)Causal QM. In Table \ref{tabop1} we suggest --- within the framework of a Bosonic system --- two important time-reversal operations $T$ and ${\cal T}$ \footnote{The time-reversal operation $T$ is supposed to exchange in- and out-states, while the anti-linear operation ${\cal T}$ is --- even for Fermionic systems --- to be identified as usual with the Hermitian conjugation, which changes causal systems into anticausal systems, and anticausal systems into causal systems. Amusingly, the author of \cite{Henry-Couannier:2004mn} working within a Hermitian framework comes to the opposite conclusion, i.e.\ that the anti-unitary time-reversal interchanges initial and final states, while the unitary time-reversal does not. The existence of two distinct time-reversal operations explains nicely, why the TCP-operations in ``Schwinger-Pauli convention'' and ``Wightman convention'' yield different results (see e.g.\ appendix A.5.\ in \cite{machet2004}!). It is tempting to define the time-reversal operation $T$ for Fermionic systems like in the case of Bosonic systems. Yet --- to our understanding --- one should additionally take into account the anticommutation properties of Fermionic operators, when interchanging initial and final states.}. As shown in Table \ref{tabop1} it is straight forward to construct from the two suggested time-reversal operations two distinct ``PT''-transformations, i.e.\ $PT$ and $P{\cal T}$. Everybody talking about ``PT''-symmetry should be well aware which of the two operations he/she is addressing! We observe also that the Hamilton operator of the Bosonic 1-dimensional (Anti)Cau-sal Harmonic Oscillator remains invariant under all introduced transformations, i.e.\ $H^{P} = H^{T} = H^{{\cal T}} = H^{PT} = H^{P{\cal T}} = H$ \footnote{As the ${\cal T}$-transformation is just the Hermitian conjugation, the ${\cal T}$-transformation symmetry of the Hamilton operator reflects just the Hermiticity of the overall Hamilton operator. Yet,  as already mentioned above, the Hamilton operator should be more accurately called {\em pseudo-Hermitian}, rather than {\em Hermitian}, as the commutation relations of the respective creation and annihilation operators are governed by an {\em indefinite metric}.} \footnote{Note that up to now we considered only Hamilton operators $H=H_C+H_A$ (and respective Lagrangeans) with $H_C=H^+_A$. In this case there appear only real energy eigenvalues and complex-valued mutually complex-conjugate pairs of energy eigenvalues. This is what we observe traditionally for PT-symmetric Hamilton operators in the phase of PT-symmetry or spontaneously broken PT-symmetry. Yet NHQT allows also models for which $H_C\not=H^+_A$. Then we are in an interesting domain, where complex energy eigenvalues do {\em not necessarily} appear as complex conjugate pairs, while analyticity, causality, Poincar\'{e}-invariance \& locality still persists, if there is at least one eigenstate serving as an asymptotic state. We shall devote future work to such models.}. The causal (anticausal) part of the Hamilton operator $H_C$ ($H_A$) of the Bosonic \mbox{1-dimensional} (Anti)Causal Harmonic Oscillator are only invariant under $P$-, $T$-, and $PT$-transformations ($H^P_C=H^T_C=H^{PT}_C=H_C$, $H^P_A=H^T_A=H^{PT}_A=H_A$), while they are not invariant under ${\cal T}$-/$P{\cal T}$-transformations ($H_C^{{\cal T}}\not= H_C$, $H_C^{P{\cal T}}\not= H_C$, $H_A^{{\cal T}}\not= H_A$, $H_A^{P{\cal T}}\not= H_A$). 
\section{Some General Aspects of (Anti)Causal Quantum Theory}
\subsection{Analyticity and the Hermiticity Content of (A)CQT}
As we mentioned in the introduction (A)CQT has to obey the postulate of non-interaction of (non-Hermitian) causal and anticausal degrees of freedom to avoid analyticity violations. This postulate defines a rather tricky interplay between the underlying Hermitian degrees of freedom, which we shall here call {\em shadow fields} according to H.P.~Stapp~\cite{Kleefeld:Nakanishi:wx,Kleefeld:Stapp:1973aa} \footnote{Note that E.C.G.\ Sudarshan \cite{Sudarshan:1974} used the term ``shadow state'' differently for states being {\em ``$\ldots$ relevant for the dynamical description but do not contribute to probability $\ldots$''} !}. We want to illustrate this decomposition into Hermitian shadow fields for the Lagrangeans of the free (anti)causal KG and Dirac theory and --- for convenience --- for the Hamilton operators of the 1-dimensional (anti)causal Bosonic and Fermionic Harmonic Oscillator, i.e.\ respectively for:
\begin{eqnarray} {\cal L} (x) & = &  
\frac{1}{2}\, \Big( (\partial \,\phi (x) )^2  - \, M^2 \, (\phi (x))^2 \, \Big)\; + \;
\frac{1}{2}\, \Big( (\partial \,\phi^+ (x) )^2  - \, M^{\ast \, 2} \,
(\phi^+ (x))^2 \Big) \; , \nonumber \\
{\cal L} (x) & = & \frac{1}{2} \,\Big( \; \overline{\psi^c} (x) \, ( \frac{1}{2} \, i \! \stackrel{\;\,\leftrightarrow}{\partial\!\!\!/} \! -  M ) \, \psi (x) \; + \; \bar{\psi} (x) \, ( \frac{1}{2} \, i \! \stackrel{\;\,\leftrightarrow}{\partial\!\!\!/} \! -  \bar{M} ) \, \psi^c (x) \; \Big) \; , \nonumber \\
H & = &
\frac{1}{2} \; \omega \, \{c^+, a\, \} + \frac{1}{2} \; \omega^\ast \, \{a^+, c\, \} \; , \quad
H \;\; = \;\; \frac{1}{2} \; \omega \, [\,d^+, b\, ] \; + \frac{1}{2} \; \omega^\ast \, [\, b^+, d\, ] \, . \qquad 
\end{eqnarray}
$\phi(x)$, $\phi^+(x)$, $\psi(x)$, $\psi^c(x)$ are decomposed in Hermitian shadow fields $\phi_{(1)}(x)$, $\phi_{(2)}(x)$, $\psi_{(1)}(x)$, $\psi_{(2)}(x)$ by  $\phi(x) = ( \phi_{(1)}(x) + i \, \phi_{(2)} (x))/\sqrt{2}$, $\phi^+(x) = ( \phi_{(1)}(x) - i \, \phi_{(2)} (x))/\sqrt{2}$, and $\psi(x) = ( \psi_{(1)}(x) + i \, \psi_{(2)} (x))/\sqrt{2}$, $\psi^c(x) = ( \psi_{(1)}(x) - i \, \psi_{(2)} (x))/\sqrt{2}$, while the respective replacement rules for the Bosonic/Fermionic Harmonic Oscillator are listed in Eq.\ (\ref{reprules1}). As a result of the replacements we obtain:
\begin{eqnarray} {\cal L} (x) & = & 
\frac{1}{2} \,\Big( (\partial \phi_{(1)} (x) )^2  - \mbox{Re}[M^2]  
(\phi_{(1)} (x))^2 \Big)  \nonumber  \\[1mm] 
 & - & \frac{1}{2} \,\Big( (\partial \phi_{(2)} (x) )^2 - \mbox{Re}[M^2]  (\phi_{(2)} (x))^2 \Big) + \mbox{Im}[M^2]  \; \phi_{(1)} (x) \,\phi_{(2)} (x) \; , \nonumber \\
{\cal L} (x) & = & \frac{1}{2} \, \overline{\psi}_{(1)} (x) \, \Big( \frac{1}{2} \, i \!\! \stackrel{\;\,\leftrightarrow}{\not\!\partial} \! -  \,\mbox{Re}[M] \Big) \psi_{(1)} (x) -  \frac{1}{2} \; \overline{\psi}_{(2)} (x) \, \Big( \frac{1}{2} \,i \!\! \stackrel{\;\,\leftrightarrow}{\not\!\partial} \! - \, \mbox{Re}[M] \Big) \psi_{(2)} (x) \nonumber  \\
 & + &   \frac{1}{2} \; \mbox{Im}[M] \; \Big( \overline{\psi}_{(2)} (x) \;\psi_{(1)} (x) +  \overline{\psi}_{(1)} (x) \;  \psi_{(2)} (x) \Big) \; , \nonumber \\ 
H & = &
\frac{1}{2}\,\mbox{Re}[\omega] \, \Big(\{ a^+_{{}_{(1)}}, a_{{}_{(1)}}\} - \{a^+_{{}_{(2)}}, a_{{}_{(2)}}\}\Big) - \mbox{Im}[\omega]  \, \big(a^+_{{}_{(1)}} a_{{}_{(2)}} + a^+_{{}_{(2)}} a_{{}_{(1)}}\big) \; , \nonumber \\
H & = &
\frac{1}{2}\,\mbox{Re}[\omega] \, \Big(\; [ \, b^+_{{}_{(1)}}, b_{{}_{(1)}}\, ] - [\, b^+_{{}_{(2)}}, b_{{}_{(2)}}\,]\; \Big) - \mbox{Im}[\omega]  \, \big(\,b^+_{{}_{(1)}} b_{{}_{(2)}} + b^+_{{}_{(2)}} b_{{}_{(1)}}\big) \; . 
\end{eqnarray}
First we observe that in all cases one shadow field has {\em positive} norm, the other has {\em negative} norm displaying the underlying {\em indefinite metric} \footnote{The normal and abnormal shadow fields appearing in (A)CQT correspond nicely to the normal and abnormal states discovered by W.\ Pauli \cite{Pauli:1956} in the Lee-model.}. Secondly we recall that shadow fields are not described by causal or anticausal propagators, but by {\em acausal} linear combinations which reduce for quasi-real masses to {\em principal value propagators} or {\em $\delta$-distributions}. Finally we see that in all cases the Lagrangeans/Hamilton operators are not diagonal in the shadow fields, while the interaction terms are proportional to the imaginary part of $M^2$, $M$ or $\omega$ being necessary to make the theory (anti)causal and analytic \footnote{If one would remove the interaction terms, one would introduce interactions between causal and anticausal fields (e.g. $\phi(x) \, \phi^+(x)$ in the KG theory, or $z z^\ast$ in the holomorphic representation of the Bosonic Harmonic Oscillator) leading not only to a {\em violation of causality}, yet also --- as described in the introduction in the context of the Lorentz non-invariance of N.\ Nakanishi's Complex-Ghost Relativistic Field Theory --- to the {\em loss of analyticity} and {\em Lorentz covariance}.}. 

According to M.\ Znojil \cite{Znojil:2001} PT-symmetric QT is characterized by {\em quasi-parities} $Q_\ell=\pm 1$, while $Q_\ell$ is introduced by the orthogonality and completeness relations $Q_\ell \left<\psi_\ell\right|H\left|\psi_{\ell^\prime}\right>= \delta_{\ell\ell^\prime}$ and $\sum_\ell \left|\psi_\ell\right> Q_\ell \left<\psi_\ell\right| = \mbox{\bf 1}$ on the right eigenstates of the Hamilton operator (See also \cite{weigert2004}!).  By decomposing the eigenstates $|n,m>$ of the (anti)causal Bosonic Harmonic oscillator given in Eq.\ (\ref{hostates1}) into Hermitian shadow states we observe that in (A)CQT the quasi-parity in the space of shadow states can take the values $\pm 1$ and $\pm i$. Hence, (A)CQT appears here clearly as a more general framework than PT-symmetric QT.
\subsection{Some Conservation Laws in (A)CQT}
For complex mass fields several important continuity equations do not hold in their traditional form \footnote{The standard continuity-like equations for the KG, Dirac and Schr\"odinger theory for a causal field with an infinitesimal negative imaginary part in the mass, i.e.\ $M=m-\frac{i}{2}\Gamma\simeq -i\,\varepsilon$, are $\partial_\mu [ \, \phi^+ (x) \; \partial^\mu \phi (x)  -  \phi (x) \; \partial^\mu \phi^+ (x) \, ]  = 2\, i \, \varepsilon \,\phi^+ (x) \, \phi (x)$, $\partial_\mu [ \, i\, \bar{\psi} (x) \; \gamma^\mu\;  \psi (x) \, ] = -\, 2\, i \, \varepsilon \,\bar{\psi} (x) \, \psi (x)$, and $i \partial_t \, [ \, \psi^+ (x) \, \psi (x) \, ] + \frac{1}{2\, m} \!\stackrel{\rightarrow}{\nabla} \!\cdot \,[ \, \psi^+ (x) \stackrel{\rightarrow}{\nabla} \psi (x) - \psi (x) \stackrel{\rightarrow}{\nabla} \psi^+ (x) \, ] = \psi^+ (x) \, [ \, V(x) - V^+(x) \,] \, \psi (x) - \frac{i\,\varepsilon}{2\, m} \,[ \, \psi^+ (x) \stackrel{\rightarrow}{\nabla}  {}^2 \, \psi (x) + \psi (x) \stackrel{\rightarrow}{\nabla}  {}^2 \, \psi^+ (x) \, ]$, respectively.
Note that all these currents are not conserved due to the finite imaginary part of the mass (here $-\varepsilon$) or non-Hermitian {\em causal potentials} $V(x)$ being Laplace-transforms of {\em causal} propagators! The breakdown of the traditional probability concept of Max Born was sketched already in Section~\ref{maxb1}. Note that even the ``massless'' causal Dirac equation $(i\!\not\! \partial + i \, \varepsilon ) \psi (x) = 0$ is not chiral invariant (See the discussion in \cite{Kleefeld:2003zj,Kleefeld:2002au}!)!}. Fortunately there can be derived \cite{Kleefeld:2003zj,Kleefeld:2002au} (See also Section~\ref{maxb1}!) new exact conservation laws respected even by fields of arbitrary complex mass.\\[1mm]
{\em Norm or probability conservation in (A)CQT:}

(Anti)causal KG/Dirac fields are decomposed into positive/negative {\em complex} frequency parts (i.e.\ $\phi (x) = \phi^{(+)}(x) + \phi^{(-)}(x)$, $\psi (x) = \psi^{(+)}(x) + \psi^{(-)}(x)$). Subtraction of equations of motion for $\phi^{(\pm)} (x)$, $\psi^{(\pm)}$, $\psi (x)$ and respective adjoints yields for the KG, Dirac and Schr\"odinger theory continuity equations $\partial_\mu [ \, \phi^{(\mp)} (x) \; \partial^\mu \phi^{(\pm)} (x)  - (\partial^\mu \phi^{(\mp)} (x))\,\phi^{(\pm)} (x) \, ]  = 0$, $\partial_\mu [ \, i\; \overline{\psi^{(\mp)c}} (x) \; \gamma^\mu\;  \psi^{(\pm)} (x) \, ] = 0$, and $i \, \partial_t \, [ \, \tilde{\psi} (x) \, \psi (x) \, ] + \frac{1}{2\, M} \!\stackrel{\rightarrow}{\nabla} \!\cdot \,[ \, \tilde{\psi} (x) \stackrel{\rightarrow}{\nabla} \psi (x) - (\stackrel{\rightarrow}{\nabla} \tilde{\psi} (x)) \,\psi (x) ] = 0$, respectively.
Note that all currents are conserved and in general non-zero, even for the neutral KG field! We conclude that the Schr\"odinger norm is $\int d^3x \, \tilde{\psi} (x) \, \psi (x)$, and {\em not} $\int d^3x \, |\psi (x)|^2\;$!\\[1mm]
{\em Charge conservation in (A)CQT:}
 
Simply charged (anti)causal systems are introduced according to the isospin concept. We define for the KG theory $\phi_\pm (x) := \big( \phi_1(x) \pm i\, \phi_2(x) \big)/\sqrt{2}$, for the Dirac theory $\psi_\pm (x) \; := \; \big(  \psi_1(x) \pm i\, \psi_2(x)\big)/\sqrt{2}$, and for the Schr\"odinger theory $\psi_\pm (x) := \big(  \psi_1(x) \pm i\, \psi_2(x)\big)/\sqrt{2}$ and $\tilde{\psi}_\pm (x) := \big(  \tilde{\psi}_1(x) \pm i\, \tilde{\psi}_2(x)\big)/\sqrt{2}$. Subtraction of causal equations of motion and the respective adjoints
 for these fields leads for the KG, Dirac and Schr\"odinger theory to the (charge conserving) continuity equations $\partial_\mu [ \, \phi_{\mp} (x) \; \partial^\mu \phi_{\pm} (x)  - (\partial^\mu \phi_{\mp} (x))\,\phi_{\pm} (x) \, ]  = 0$, $\partial_\mu [ \, i\; \overline{\psi^{\,c}_{\mp}} (x) \; \gamma^\mu\;  \psi_{\pm} (x) \, ] = 0$, and $i \partial_t \, [ \, \tilde{\psi}_\mp (x) \, \psi_\pm (x) \, ] + \frac{1}{2\, M} \stackrel{\rightarrow}{\nabla} \!\cdot \,[ \, \tilde{\psi}_\mp (x) \stackrel{\rightarrow}{\nabla} \psi_\pm (x) - (\stackrel{\rightarrow}{\nabla} \tilde{\psi}_\mp (x)) \,\psi_\pm (x) \, ] = 0$, respectively \footnote{Note that currents and charges vanish for neutral KG and Dirac fields (The neutral theory follows by setting either $\phi_1(x)$ ($\psi_1(x)$, $\tilde{\psi}_1(x)$) or $\phi_2(x)$ ($\psi_2(x)$, $\tilde{\psi}_2(x)$) equal to zero!). The reason is a cancellation of underlying norm currents! In Schr\"odinger theory being just a theory first order in time this cancellation does not occur. Note that the neutral Dirac field does {\em not} admit any {\em Abelian} gauge couplings due to transposition properties $[\,\overline{\psi^{\,c}} (x) \not\!\!A(x) \, \psi (x)\,]^T = \! - \,\overline{\psi^{\,c}} (x) \not\!\!A(x) \, \psi (x)$, $[\,\overline{\psi^{\,c}} (x) \,\sigma^{\mu\nu} \, F_{\mu\nu}(x) \, \psi (x)\,]^T = - \,\overline{\psi^{\,c}} (x) \,\sigma^{\mu\nu} \, F_{\mu\nu}(x) \, \psi (x) $.}.
The concept of local (non)Abelian gauge invariance in (A)CQT is sketched in the footnote \footnote{The Dirac Lagrangean with minimally coupled (non)Abelian gauge fields is given by the expression ${\cal L} (x) = \overline{\psi_+^c} (x) \, ( \frac{1}{2} \,\, i \!\! \!\stackrel{\;\,\leftrightarrow}{\not\!\partial} + g \not\!\!A (x) -  M ) \, \psi_- (x) + \overline{\psi}_- (x) \, ( \frac{1}{2} \,\, i \! \!\!\stackrel{\;\,\leftrightarrow}{\not\!\partial} \!\! + g^\ast \gamma^{\,\mu} A^+_{\mu} (x) -  \bar{M} ) \, \psi_+^c (x)$ (with $\psi_\pm (x) := (\psi_1(x)\pm \, i\psi_2(x))/\sqrt{2}$). 
It is invariant under the local gauge transformations $g \not\!\!\!A^{\,\prime} = g \not\!\!\!A + [\not\!\!\partial , \Lambda (x)]$, $\psi^{\,\prime}_- (x) = \exp(+i \, \Lambda (x)) \, \psi_- (x)$, and $\psi^{\,\prime}_+ (x) = \exp(- i \, (\Lambda (x))^T) \, \psi_+ (x)$. The non-Abelian yields $A_\mu (x) =  A_\mu^{a} (x) \lambda^a/2$ and $\Lambda (x) =  \Lambda^{a} (x) \lambda^a/2$.
{\em Non-Abelian} gauge fields admit minimal coupling even to neutral Fermions, {\em if} $[A^{\mu}(x)]^T=-A^{\mu}(x)$, as $[\,\overline{\psi^{\,c}} (x) \not\!\! A(x) \, \psi (x)\,]^T = + \,\overline{\psi^{\,c}} (x) \not\!\! A(x) \, \psi (x)$ and $[\,\overline{\psi^{\,c}} (x) \,\sigma^{\mu\nu} \, F_{\mu\nu}(x) \, \psi (x)\,]^T = + \,\overline{\psi^{\,c}} (x) \,\sigma^{\mu\nu} \, F_{\mu\nu}(x) \, \psi (x) $.}.

\subsection{Conjugate $T$-Matrix $\overline{T}_{fi}$, Dual Vacuum,  Complex Transition Probabilities and Complex (Anti)Causal Cross Sections}
As $|\psi(x)|^2$ is not a probability density in (anti)causal Schr\"odinger theory \cite{Kleefeld:2002au}, $|T_{fi}|^2$ is {\em not} to be interpreted as a {\em transition probability} in (anti)causal scattering theory! In (anti)causal scattering theory we have instead to consider a quantity $\overline{T}_{fi} \,T_{fi}$, where $\overline{T}_{fi}$ ($\not= T^+_{fi}$) is called the {\em conjugate $T$-matrix}. The construction of $\overline{T}_{fi}$ has been addressed in \cite{Kleefeld:2003zj,Kleefeld:2003dx}. First we denote the causal $T$-matrix $T_{fi}$ in the interaction picture:
\begin{eqnarray} \lefteqn{(2\pi)^4 \; \delta^4 (P_f - P_i)\;\; i \; T_{fi} \; =} \nonumber \\[1mm]
 & = & \left<\!\left<0\right|\right. {\cal A}(\vec{p}^{\,\,\prime}_{\,N_f}) \ldots {\cal A}(\vec{p}^{\,\,\prime}_{\,1}) \; T [ \, \exp(+\,i \, S_{int}) - 1 \, ]\; ({\cal C}(\vec{p}_{\,1}))^+ \ldots ({\cal C}(\vec{p}_{\,N_i}))^+ \left|0\right>_c \nonumber \\[1mm]
 & \equiv & \left<\!\left<\psi_f\right|\right. T [ \, \exp(i \, S_{int}) - 1 \, ] \left|\psi_i\right>
\end{eqnarray}
with ${\cal A}(\vec{p}^{\,\,\prime}_{j}) \in \{ a (\vec{p}^{\,\,\prime}_{j}), b (\vec{p}^{\,\,\prime}_{j})\}$ and  ${\cal C}(\vec{p}_{j})\in \{ c (\vec{p}_{j}), d (\vec{p}_{j})\}$. Call $N_F$ the overall number of Fermionic operators in the intitial and final state. Then we obtain for $\overline{T}_{fi}$:
\begin{eqnarray} \lefteqn{(2\pi)^4 \; \delta^4 (P_f - P_i)\; \; (-i) \;\, \overline{T}_{if} \quad =} \nonumber \\
 & = & \left<\!\left<\bar{0}\right|\right. \Big( {\cal A}(\vec{p}_{\,N_i}) \ldots {\cal A}(\vec{p}_{\,1}) \; T [ \, \exp(-\,i \, S_{int}) - 1 \, ]\; ({\cal C}(\vec{p}^{\,\,\prime}_{\,1}))^+ \ldots ({\cal C}(\vec{p}^{\,\,\prime}_{\,N_f}))^+ \Big)^T \left|\bar{0}\right>_c \nonumber \\
 & \stackrel{!}{=} & (-1)^{N_F (N_F-1)/2} \times \nonumber \\
 & & \times \left<\!\left<\bar{0}\right|\right. ({\cal C}(\vec{p}^{\,\,\prime}_{\,N_f}))^+ \ldots ({\cal C}(\vec{p}^{\,\,\prime}_{\,1}))^+ \; \Big( T \,[ \, \exp(-\,i \, S_{int}) - 1 \, ]\Big)^T {\cal A}(\vec{p}_{\,1}) \ldots {\cal A}(\vec{p}_{\,N_i}) \left|\bar{0}\right>_c \nonumber \\
 & = & (-1)^{N_F (N_F-1)/2} \times \nonumber \\
 & & \times  \left<\!\left<\bar{0}\right|\right. ({\cal C}(\vec{p}^{\,\,\prime}_{\,N_f}))^+ \ldots ({\cal C}(\vec{p}^{\,\,\prime}_{\,1}))^+ \; \overline{T} \,[ \, \exp(-\,i \, S^{\,T}_{int}) - 1 \, ]\; {\cal A}(\vec{p}_{\,1}) \ldots {\cal A}(\vec{p}_{\,N_i}) \left|\bar{0}\right>_c \nonumber \\
 & \stackrel{!}{=} & (-1)^{N_F (N_F-1)/2} \times \nonumber \\
 & & \times  \left<\!\left<\bar{0}\right|\right. ({\cal C}(\vec{p}^{\,\,\prime}_{\,N_f}))^+ \ldots ({\cal C}(\vec{p}^{\,\,\prime}_{\,1}))^+ \, \overline{T} \,[ \, \exp(-\,i \, S_{int}) - 1 \, ]\, {\cal A}(\vec{p}_{\,1}) \ldots {\cal A}(\vec{p}_{\,N_i}) \left|\bar{0}\right>_c \label{contmat1}
\end{eqnarray}
with ${\cal A}(\vec{p}_{j}) \in \{ a (\vec{p}_{j}), b (\vec{p}_{j})\}$ and  ${\cal C}(\vec{p}^{\,\,\prime}_{j})\in \{ c (\vec{p}^{\,\,\prime}_{j}), d (\vec{p}^{\,\,\prime}_{\,j})\}$ and
$\left|\bar{0}\right>$ (and $\left<\!\left<\bar{0}\right|\right.$) being the {\em dual vacuum} annihilating creation operators and creating annihilation operators (See also Section \ref{repfree1}!).
Above we used the useful identity  
$\big( T\, \big[ \,{\cal O}(x_1) \ldots  {\cal O}(x_n) \, \big] \big)^T \; = \; \overline{T}\, \big[ \,({\cal O}(x_1))^T \ldots  ({\cal O}(x_n))^T \, \big]$ holding for Bosonic and Fermionic operators. Eq.\ (\ref{contmat1}) may be rewritten in the following way:
\begin{eqnarray}
\lefteqn{(2\pi)^4 \; \delta^4 (P_f - P_i)\; \; (-i) \;\, \overline{T}_{if} \quad =} \nonumber \\
 & = & \left|\bar{0}\right>^T \, {\cal A}(\vec{p}_{\,N_i}) \ldots {\cal A}(\vec{p}_{\,1}) \; T [ \, \exp(-\,i \, S_{int}) - 1 \, ]\; ({\cal C}(\vec{p}^{\,\,\prime}_{\,1}))^+ \ldots ({\cal C}(\vec{p}^{\,\,\prime}_{\,N_f}))^+ \, \big(\left<\!\left<\bar{0}\right|\right.\big)^T \nonumber \\
 & \equiv & \left|\right.\!\overline{\psi}_i\!\left.\right>^T \, T [ \, \exp(- i \, S_{int}) - 1 \, ]  \left(\left<\!\left<\right.\right.\!\overline{\psi}_f\!\!\left.\left.\right|\right.\right)^T \; .
\end{eqnarray}
The (eventually complex) transition probability for a causal process is \footnote{Hence the (complex) ``probability'' of a state $\left|\psi\right>$ to be in a state $\left|Y\right>$ appears to be $\left|\right.\!\overline{\psi}\!\left.\right>^T \! \left<\!\left<\right.\right.\!\overline{Y}\!\!\left.\left.\right|\right.^T \left<\!\left<Y\right|\right.\!\!\left.\psi\right>=
\left|\right.\!\overline{Y}\!\left.\right>^T \! \left<\!\left<\right.\right.\!\overline{\psi}\!\!\left.\left.\right|\right.^T
\left<\!\left<\psi\right|\right.\!\!\left.Y\right>$. For a further short discussion see \cite{Kleefeld:2003zj}.} \footnote{If $\overline{T}_{fi} \,T_{fi}$ is complex for a causal process, then also the respective causal cross section will be complex. Only if the underlying theory represented by a causal Lagrangean or causal Hamiltonian is probability conserving, i.e.\ non-absorptive, $\overline{T}_{fi} \,T_{fi}$ and therefore also the causal cross section will be quasi-real, i.e.\ infinitesimally close to a real number.
 Hence, for selective (so called ``inelastic'') causal processes in probability non-conserving theories (e.g.\ open quantum systems) the respective causal cross sections will develop a finite imaginary part. Within particle physics this new feature complements in a beautiful manner, what is well understood for a long time in theoretical optics, i.e.\ that the imaginary part of the refractive index of a material is related to its absorption coefficient. Similar arguments hold for anticausal processes and respective anticausal cross sections. A temptative interpretation of complex cross sections was provided by T.\ Berggren \cite{berggren1970}. Furthermore, he suggested to study {\em experimentally observable} effects to measurable lineshapes in inelastic scattering experiments due to complex cross sections. Inversely we suggest that a reality constraint on causal cross sections will yield important theoretical constraints on (mass and coupling) parameters of Lagrangeans used to describe elastic scattering processes.}:
\begin{eqnarray} \lefteqn{\overline{T}_{fi} \,T_{fi} \; = \; \left|\right.\!\overline{\psi}_i\!\left.\right>^T\, T [ \, \exp(- i \, S_{int}) - 1 \, ]  \left(\left<\!\left<\right.\right.\!\overline{\psi}_f\!\!\left.\left.\right|\right.\right)^T\! \left<\!\left<\psi_f\right|\right. T [ \, \exp(i \, S_{int}) - 1 \, ] \left|\psi_i\right> } \nonumber \\
 & = & (-1)^{N_F(N_F-1)/2} \,\times \nonumber \\
 & & \times  \left<\!\left<\right.\right.\!\overline{\psi}_f\!\!\left.\left.\right|\right. \!\big( T [ \, \exp(- i \, S_{int}) - 1 \, ]\big)^T \! \left|\right.\!\overline{\psi}_i\!\left.\right> \left<\!\left<\psi_f\right|\right. T [ \, \exp(i \, S_{int}) - 1 \, ] \left|\psi_i\right> \; . 
\end{eqnarray}

\section{Applications}
\subsection{The ``Shifted'' (Anti)Causal Harmonic Oscillator}
In 1997 C.M.\ Bender \& S.\ Boettcher \cite{Bender:1998ke} (See also M.\ Znojil \cite{Znojil:1999qt,Znojil:2001}) used the non-Hermitian Hamilton operator $H=p^2 + x^2 + i\,x= p^2 + (x+i/2)^2 +1/4$ obtained from a Harmonic Oscillator shifted to a complex space point $x=-i/2$ as an example to show that its spectrum $E_n=(2\,n+ 1) + 1/4 = 2\,n + 5/4$ can be indeed real due to the underlying ``PT-symmetry''. Noting that in this work there is seemingly not made any distinction between the real variable $x$ and the complex variables $z$ and $z^\ast$ appearing in the holomorphic representation, we want to illustrate here the explicit formalism for a respective shift of the (Anti)Causal Bosonic Harmonic Oscillator within (A)CQT. The Hamilton operator of the ``unshifted'' (Anti)Causal Bosonic Harmonic Oscillator in holomorphic representation and in ``representation free'' form are given by:
\begin{eqnarray}
H(z,z^\ast) & = & -\, \frac{1}{2\, M} \,\frac{d^2}{dz^2} + \frac{1}{2} \; M \, \omega^2 \, z^2 -\, \frac{1}{2\, M^\ast} \, \frac{d^2}{dz^{\ast 2}} + \frac{1}{2} \; M^\ast \, \omega^{\ast 2} \, z^{\ast 2} \; , \nonumber \\
H & = &
\frac{1}{2} \; \omega \, \{c^+, a\, \}
+ \frac{1}{2} \; \omega^\ast \, \{a^+, c\, \} \; .
\end{eqnarray}
Without changing its spectrum the Hamilton operator is shifted from $(z,z^\ast)$ to \mbox{$(z+\alpha,z^\ast+\beta^\ast)$} by applying the following operator, which is denoted in the holomorphic representation as $U_{z,z^\ast}(\alpha,\beta^\ast)$ and in the representation free case as $U[\alpha,\beta^\ast]$:
\begin{eqnarray} U_{z,z^\ast}(\alpha,\beta^\ast) & = & \exp\Big( i \, ( \alpha \, p + \beta^\ast \, p^\ast ) \Big) \;\; = \;\; \exp\Big( + \alpha \, \frac{d}{dz} + \beta^\ast \, \frac{d}{dz^\ast} \Big) \; ,  \nonumber \\ 
U[\alpha,\beta^\ast] & = & \exp\Big( i\,\alpha \, \sqrt{\frac{M\,\omega}{2}} \; (a+c^+) + i\, \beta^\ast \, \sqrt{\frac{M^\ast\,\omega^\ast}{2}} \; (c+a^+) \Big) \; .
\end{eqnarray}
The respective inverse operators are given by $U^{-1}_{z,z^\ast}(\alpha,\beta^\ast)\equiv U_{z,z^\ast}(-\alpha,-\beta^\ast)$ and $U^{-1}[\alpha,\beta^\ast]\equiv U[-\alpha,-\beta^\ast]$. The ``shifted'' Hamilton operator obtained by a standard equivalence transform $H(z+\alpha,z^\ast+\beta^\ast) = U_{z,z^\ast}(\alpha,\beta^\ast) \; H(z,z^\ast) \; U^{-1}_{z,z^\ast}(\alpha,\beta^\ast)$ and $H[\alpha,\beta^\ast]= U[\alpha,\beta^\ast] \; H \; U^{-1}[\alpha,\beta^\ast]$ is respectively:
\begin{eqnarray} H(z+\alpha,z^\ast+\beta^\ast) & = & H(z,z^\ast) + M \, \omega^2 \, \alpha \left(z+\frac{\alpha}{2}\right)  \; +  M^\ast \, \omega^{\ast \,2} \, \beta^\ast \left(z^\ast + \frac{\beta^\ast}{2}\right) \; , \nonumber \\
H[\alpha,\beta^\ast] & = & H + M \, \omega^2 \, \alpha \left(\frac{i}{\sqrt{2\,M\, \omega}} \; (a - c^+)+\frac{\alpha}{2}\right)  \nonumber \\
 & & + \,  M^\ast \, \omega^{\ast \,2} \, \beta^\ast \left(\frac{i}{\sqrt{2\,M^\ast\, \omega^\ast}} \; (c - a^+) + \frac{\beta^\ast}{2}\right) \, .
\end{eqnarray}
The ``shifted'' Hamilton operator will be $P{\cal T}$-symmetric (i.e.\ $H(z+\alpha,z^\ast+\beta^\ast)^{P{\cal T}}=H(z+\alpha,z^\ast+\beta^\ast)$) for $\alpha=-\beta \Leftrightarrow \alpha^\ast=-\beta^\ast$. It will be Hermitian/${\cal T}$-symmetric (i.e.\ $H(z+\alpha,z^\ast+\beta^\ast)^{{\cal T}}=H(z+\alpha,z^\ast+\beta^\ast)$) for $\alpha=\beta \Leftrightarrow \alpha^\ast=\beta^\ast$. Bender's example ``$H=p^2 + x^2 + i\,x\,$'' is essentially obtained by $H(z+i\,\gamma,z^\ast+i\,\gamma^\ast)$ with $\gamma = +1/2$. For $\omega\not=\omega^\ast$ --- even being strictly $P{\cal T}$-symmetric --- only some part of its spectrum $E_{n,m}=\omega \,(n+\frac{1}{2})+\omega^\ast (m+\frac{1}{2})$ with $n,m\in{\rm I\!N}$${}_0$ is real, namely if $n=m$.

\subsection{Non-Hermitian Supersymmetry}
Non-Hermitian Hamilton supersymmetric operators with PT-symmetry have been investigated already for quite some time (See e.g.\ \cite{Znojil:2000nh,Znojil:2000fr,Dorey:2001hi,Znojil:2002wi,Znojil:2002pq,Mostafazadeh:2004ph,caliceti2004b}!). We want here to go one step further and
consider the non-PT-symmetric Hamilton operator of an (Anti)Causal Supersymmetric Harmonic Oscillator, being the sum of a causal Bosonic and
Fermionic Harmonic Oscillator with equal {\em complex} frequency $\omega_C$, and an anticausal Bosonic and
Fermionic Harmonic Oscillator with equal {\em complex} frequency $\omega_A$:
\begin{eqnarray}
H & = &
\frac{1}{2} \; \omega_C \, \{c^+, a\, \}
+\frac{1}{2} \; \omega_C \, [\,d^+, b\, ]
+ \frac{1}{2} \; \omega_A^\ast \, \{a^+, c\, \}
+ \frac{1}{2} \; \omega_A^\ast \, [\, b^+, d\, ] \nonumber \\
& = &
\omega_C \, \Big(\,c^+ a+\frac{1}{2}\, \Big)
+ \omega_C \, \Big(\,d^+ b\,-\frac{1}{2} \,\Big)
+\omega_A^\ast \Big(\,a^+ c+\frac{1}{2}\, \Big)
+ \omega_A^\ast \Big(\,b^+ d\,-\frac{1}{2} \,\Big)
 \nonumber \\[1mm]
& = &
\omega_C \; (c^+ a+ d^+ b )
+\omega_A^\ast \, (a^+ c+ b^+ d ) \; .
\end{eqnarray}
As usual in a supersymmetric system the positive and negative
contributions to the Bosonic and Fermionic vacuum energy, respectively,
cancel.
The Hamilton operator is easily diagonalized by the {\em normal} right eigenstates $\big| n_B,n_F;\bar{n}_B,\bar{n}_F\big>$ or {\em normal} left eigenstates $\big<\!\big< n_B,n_F;\bar{n}_B,\bar{n}_F\big|$ given by
\begin{eqnarray} \big| n_B,n_F;\bar{n}_B,\bar{n}_F \big> & = & \frac{1}{\sqrt{n_B!\,\bar{n}_B!}} \,\, (c^+)^{n_B} (d^+)^{n_F} (a^+)^{\bar{n}_B} (b^+)^{\bar{n}_F}\big|0\big> \; , \nonumber \\
\big<\!\big< n_B,n_F;\bar{n}_B,\bar{n}_F \big| & = & \frac{1}{\sqrt{n_B!\,\bar{n}_B!}} \,\, \big<\!\big< 0 \big| \; d^{\,\bar{n}_F} \; c^{\bar{n}_B} \; b^{n_F} \; a^{n_B} \; , 
\end{eqnarray}
(with $n_B$, $\bar{n}_B \in {\rm I\!N}$${}_0$ and  $n_F$, $\bar{n}_F \in \{0,1\}$) yielding the respective energy eigenvalues $E_{n_B,n_F;\bar{n}_B,\bar{n}_F} = \omega_C (n_B+n_F) + \omega_A^\ast (\bar{n}_B+\bar{n}_F)$.
Now we define supercharges $Q_\pm$ and respective
Hermitian conjugate supercharges $Q^+_\pm$ by:
\begin{equation} Q_+ = a \,d^+\; , \quad Q_+^+ = d \,a^+ \; , \quad Q_- = c^+\,b \; , \quad Q_-^+ = b^+\,c \; .
\end{equation}
The supercharges $Q_+$ and $Q_-$ (and $Q^+_+$ and $Q^+_-$) are nilpotent, yet {\em not} related by Hermitian conjugation, i.e.:
\begin{equation} Q_\pm^2 \; = \; \left( Q_\pm^+ \right)^2 = 0 \quad , \quad \left( Q_\pm
\right)^+ \; \not= \;  Q_\mp \; .
\end{equation}
The Hamilton operator can be expressed in terms of supercharges as follows:
\begin{eqnarray}
H & = &
\omega_C \; \{ Q_+ , Q_-\}
+\omega_A^\ast \, \{ Q^+_- , Q^+_+\} \; .
\end{eqnarray}
Application of supercharges interrelates as usual distinct eigenstates of the Hamilton operator belonging to the same eigenvalue. For {\em normal} right eigenvectors we have:
\begin{eqnarray} Q_+ \; \big| n_B,n_F;\bar{n}_B,\bar{n}_F\big> & \propto & \big| n_B - 1,n_F + 1;\bar{n}_B,\bar{n}_F\big> \; , \nonumber \\
Q_- \; \big| n_B,n_F;\bar{n}_B,\bar{n}_F\big> & \propto & \big| n_B + 1,n_F - 1;\bar{n}_B,\bar{n}_F\big> \; , \nonumber \\
Q^+_+ \; \big| n_B,n_F;\bar{n}_B,\bar{n}_F\big> & \propto & \big| n_B,n_F;\bar{n}_B+1,\bar{n}_F-1\big> \; , \nonumber \\
Q^+_- \; \big| n_B,n_F;\bar{n}_B,\bar{n}_F\big> & \propto & \big| n_B,n_F;\bar{n}_B-1,\bar{n}_F+1\big> \; .
\end{eqnarray}
In order to compare this non-Hermitian supersymmetric Harmonic Oscillator with its Hermitian counterpart one has to decompose creation and annihilation operators of this oscillator into their Hermitian (``shadow'') components \footnote{The decomposition in Hermitian (``shadow'') components is performed as follows: 
\begin{eqnarray}
a & = & \frac{1}{\sqrt{2}} \; (a_{{}_{(1)}}+i \,a_{{}_{(2)}}) \; , \quad a^+ \; = \; \frac{1}{\sqrt{2}} \; (a^+_{{}_{(1)}}-i \,a^+_{{}_{(2)}}) \; , \nonumber \\
c & = & \frac{1}{\sqrt{2}} \; (a_{{}_{(1)}}-i \,a_{{}_{(2)}}) \; , \quad c^+ \; = \; \frac{1}{\sqrt{2}} \; (a^+_{{}_{(1)}}+i \,a^+_{{}_{(2)}}) \; ,\nonumber \\
b & = & \frac{1}{\sqrt{2}} \; (b_{{}_{(1)}}+i \,b_{{}_{(2)}}) \; , \quad \; b^+ \; = \; \frac{1}{\sqrt{2}} \; (b^+_{{}_{(1)}}-i \,b^+_{{}_{(2)}})\; ,\nonumber \\
d & = & \frac{1}{\sqrt{2}} \; (b_{{}_{(1)}}-i \,b_{{}_{(2)}}) \; , \quad \, d^+ \; = \; \frac{1}{\sqrt{2}} \; (b^+_{{}_{(1)}}+i \,b^+_{{}_{(2)}}) \; . \label{reprules1}
\end{eqnarray}
The creation and annihilation operators of the Hermitian shadow fields obey the
following (anti)commutation relations:
\begin{eqnarray} \left( \begin{array}{cc} {[a_{{}_{(1)}},a^+_{{}_{(1)}}]} & {[a_{{}_{(1)}},a^+_{{}_{(2)}}]} \\
{[a_{{}_{(2)}},a^+_{{}_{(1)}}]} & {[a_{{}_{(2)}},a^+_{{}_{(2)}}]} \end{array}\right) & = & \left( \begin{array}{cc} {\{b_{{}_{(1)}},b^+_{{}_{(1)}}\}} & {\{b_{{}_{(1)}},b^+_{{}_{(2)}}\}} \\
{\{b_{{}_{(2)}},b^+_{{}_{(1)}}\}} & {\{b_{{}_{(2)}},b^+_{{}_{(2)}}\}} \end{array}\right) \; = \; \left( \begin{array}{cc} 1 & 0 \\
0 & -1 \end{array}\right)  \; ,
\nonumber \\[1mm]
 \left( \begin{array}{cc} {[a_{{}_{(1)}},a_{{}_{(1)}}]} & {[a_{{}_{(1)}},a_{{}_{(2)}}]} \\
{[a_{{}_{(2)}},a_{{}_{(1)}}]} & {[a_{{}_{(2)}},a_{{}_{(2)}}]} \end{array}\right) & = & \left( \begin{array}{cc} {[a^+_{{}_{(1)}},a^+_{{}_{(1)}}]} & {[a^+_{{}_{(1)}},a^+_{{}_{(2)}}]} \\
{[a^+_{{}_{(2)}},a^+_{{}_{(1)}}]} & {[a^+_{{}_{(2)}},a^+_{{}_{(2)}}]} \end{array}\right) \;\;\, = \; \left( \begin{array}{cc} 0 & 0 \\
0 & 0 \end{array}\right) \; , \nonumber \\[1mm]
 \left( \begin{array}{cc} {\{b_{{}_{(1)}},b_{{}_{(1)}}\}} & {\{b_{{}_{(1)}},b_{{}_{(2)}}\}} \\
{\{b_{{}_{(2)}},b_{{}_{(1)}}\}} & {\{b_{{}_{(2)}},b_{{}_{(2)}}\}} \end{array}\right) & = & \left( \begin{array}{cc} {\{b^+_{{}_{(1)}},b^+_{{}_{(1)}}\}} & {\{b^+_{{}_{(1)}},b^+_{{}_{(2)}}\}} \\
{\{b^+_{{}_{(2)}},b^+_{{}_{(1)}}\}} & {\{b^+_{{}_{(2)}},b^+_{{}_{(2)}}\}} \end{array}\right) \; = \; \left( \begin{array}{cc} 0 & 0 \\
0 & 0 \end{array}\right) . \qquad
\end{eqnarray}
As usually Fermionic and Bosonic field operators commute.}.
The Hamilton operator of the (Anti)Causal Supersymmetric Harmonic Oscillator in
terms of creation and annihilation operators of the Hermitian shadow fields is
given by:
\begin{eqnarray}
H & = &
\frac{\omega_C+\omega^\ast_A}{2} \, (a^+_{{}_{(1)}} a_{{}_{(1)}} - a^+_{{}_{(2)}} a_{{}_{(2)}} + b^+_{{}_{(1)}} b_{{}_{(1)}} - b^+_{{}_{(2)}}
b_{{}_{(2)}})\nonumber \\
& - &  \frac{\omega_C-\omega^\ast_A}{2\, i} \, (a^+_{{}_{(1)}} a_{{}_{(2)}} + a^+_{{}_{(2)}} a_{{}_{(1)}} + b^+_{{}_{(1)}} b_{{}_{(2)}} + b^+_{{}_{(2)}}
b_{{}_{(1)}}) \; .
\end{eqnarray}
This expression reflects most transparently the tricky interplay of normal/abnormal Bosonic/Fermionic shadow fields, when providing us with supersymmetry even for arbitrary complex-valued oscillator frequencies $\omega_C$ and $\omega_A$. The PT-symmetric Hamilton operator is obtained by setting $\omega\equiv\omega_C=\omega_A$ yielding $(\omega_C+\omega^\ast_A)/2=$Re$[\omega]$ and $(\omega_C-\omega^\ast_A)/(2\,i)=$Im$[\omega]$. The limit Im$[\omega]=0$ was considered recently in \cite{Mostafazadeh:2004ph}, where the abnormal Fermions $b_{{}_{(2)}}$ and $b^+_{{}_{(2)}}$ are called {\em Phermions}~\footnote{Note that the author of \cite{Mostafazadeh:2004ph} uses the notation ${\{b_{{}_{(2)}},b^{\,\sharp}_{{}_{(2)}}\}}=-1$ rather than ${\{b_{{}_{(2)}},b^+_{{}_{(2)}}\}}=-1$, as he believes that ${\{b_{{}_{(2)}},b^+_{{}_{(2)}}\}}$ should be a positive operator. Instead, we follow here \mbox{L.K.~Pandit}~\cite{pandit1959} who understands ---  as discussed in the introduction --- that Hermitian conjugation cannot be defined without specifying the metric!}.
\subsection{Towards a Theory of Strong Interactions without Gluons}
Is is common belief that the theory of strong interaction is described by QCD \cite{shi2001a}~\footnote{This belief is strongly supported by experiments probing the perturbative regime of QCD and displaying there successfully a key feature of QCD, i.e.\ {\em asymptotic freedom}. Furthermore, as discussed in \cite{Gross:1973ju}, it had not only been shown \cite{Gross:1973id} that non-Abelian gauge theories are asymptotically free, yet also ``proved'' by S.\ Coleman \& D.J.\ Gross \cite{Coleman:1973sx} that {\em ``$\ldots$ no renormalizable field theory without non-Abelian gauge fields can be asymptotically free $\ldots$''}. The absence of asymptotic freedom in QED had been proved earlier by M.\ Gell-Mann \& F.\ Low \cite {Gell-Mann:1954fq} and ``shown'' for (pseudo)scalar-Fermion theories involving one coupling constant by A.~Zee \cite{Zee:1973gn}.}, i.e.\ gluons and (anti)quarks. The clear experimental evidence allowing such a belief (in particular with respect an eventual existence of gluons) is still lacking \footnote{Experimental observations e.g.\ of multijet events in high-energy physics experiments cannot yet convincingly be related to gluons, as they may be also well due to different partonic structures.}. On the theoretical side there seem to be now even strong arguments against the existence of gluons due to calculations by R.\ Alkofer et al. \cite{Alkofer:2003jk} (See also \cite{Cucchieri:2004mf}!) who observe that {\em spectral functions of gluons violate positity} \footnote{Just a warning: it may well be that these positivity violations are just an artefact due to some problems with (non-Abelian) Gupta-Bleuler projection on physical states in the context of truncated systems of Dyson-Schwinger equations or on the lattice.}, while the spectral functions of quarks do {\em not}! At the first sight there does not seem to be a substitute for gluons at hand, which allows to achieve asymptotic freedom without getting in conflict with the statements by Coleman, Gross, and Zee, as asymptotically free theories with e.g.\ scalars and Fermions are required to be rather more \cite{Vaughn:1981qi} than less \cite{HariDass:1999mj} {\em exotic}. Mysteriously, the Quark-Level Linear Sigma Model \cite{Delbourgo:1993dk} (QLL$\sigma$M) has been a rather successful theory to describe various experimental facts involving hadronic physics at low and intermediate energy (See e.g.\ \cite{Kleefeld:2001ds} and references therein!). Guided by this observation we ``mapped'' \cite{Kleefeld:2002au} in 2002 by a simplistic argument the Lagrangean of QCD (Strong coupling $g$\,!) into a Lagrangean of a QLL$\sigma$M which is supposed to describe quark-quark scattering equally well {\em at high energies}. In {\em unitary gauge} we obtained the following QLL$\sigma$M Lagrangean:
\begin{eqnarray} {\cal L} & = & \overline{q^c_+} \; \Big( \,\frac{i}{2} \!\! \stackrel{\;\,\leftrightarrow}{\partial\!\!\!/} \!\! -  M + g \, \sqrt{\frac{N_F}{N_c}} \big[\sqrt{2}\; i \,(s_s \, S + s_p \,i\,P\,\gamma_5) + \frac{1}{\sqrt{2}} (s_v \not\!{V} + s_y \not\!{Y}\gamma_5)  \big] \Big) \, q_- \nonumber \\
 & & + \, \frac{1}{2} \,\mbox{tr} \Big[ \Big( \partial^\mu S - i\,g \, \sqrt{\frac{N_F}{2\, N_c}} \big(s_v [ V^\mu , S]  - i\, s_s s_p s_y  \{Y^\mu, P\} \big) \Big)^2  \Big] - \frac{1}{2} \, M^2 \,\mbox{tr} \big[ S^2 \big] \nonumber \\
 & & + \, \frac{1}{2} \,\mbox{tr} \Big[ \Big( \partial^\mu P - i\,g \, \sqrt{\frac{N_F}{2\, N_c}} \big(s_v[ V^\mu , P]  + i\, s_s s_p s_y \{Y^\mu, S\} \big) \Big)^2  \Big] - \frac{1}{2} \, M^2 \,\mbox{tr} \big[ P^2 \big] \nonumber \\
 & & - \, \frac{1}{8} \; \mbox{tr} \big[ (V^{\mu\nu}_+)^2 \big] + \frac{1}{4} \; M^2 \, \mbox{tr} \big[ \big(V^\mu_+\big)^2 \big] \; - \, \frac{1}{8} \; \mbox{tr} \big[ (V^{\mu\nu}_-)^2 \big] + \frac{1}{4} \; M^2 \, \mbox{tr} \big[ \big(V^\mu_-\big)^2 \big] \nonumber \\
 & & -\, \frac{\lambda}{2} \; \mbox{tr} \big[ \big((S+  i\, P)(S-  i\, P) \big)^2 \big] + \mbox{gauge fixing terms/ghosts} + \mbox{h.c.}, \label{linsig1}
\end{eqnarray}
with $s_s$, $s_p$, $s_v$, $s_y\in\{+1,-1\}$, $V^\mu_\pm = V^\mu \pm Y^\mu$, $V^{\mu\nu}_\pm = [D^\mu_\pm,D^\nu_\pm]/(-i\, g\,\sqrt{N_F/(2\,N_c)}\,)$, $D^\mu_\pm = \partial^\mu -i\, g\,\sqrt{N_F/(2\,N_c)}\, V^\mu_\pm$, tr$=$``flavour-trace'', and $S$, $P$, $V^\mu$, $Y^\mu$ are scalar, pseudoscalar, vector and axial vector $N_F\times N_F$ meson field matrices in flavour space, respectively, while $\lambda = (g \, \sqrt{N_F/N_c}\,)^2$. 
This unexpected result is a {\em non-Hermitian} QLL$\sigma$M Lagrangean, which is {\em asymptotically free} due to a purely {\em imaginary} Yukawa coupling, the PT-symmetry of which suggests in correspondence to C.M.\ Bender's ``physical'' $i\,\phi^3$-theory \cite{Bender:1998ke,Bender:2004by} a {\em real spectrum} of the respective Hamilton operator! Additionally we observe that all vertices are accompanied by weightfactors completely in accordance with the $1/\sqrt{N_c}$ expansion \cite{'tHooft:1974hx}, while the $V$ and $Y$ mesons couple to the $S$ and $P$ mesons in a similar manner as in an 1-loop effective action of the Gauged Linear Sigma Model in \cite{alk1} or the Extended Chiral Quark Model in \cite{andrianov1} obtained by a tedious Bosonization from the QCD Lagrangean. Furthermore we observe that the obtained Lagrangean provides strictly Sakurai's {\em vector meson universality} required in \cite{Djukanovic:2004mm} by renormalization constraints. Hence, the obtained Lagrangean is a theoretically very admissable Lagrangean for the description of strong interactions, which may be used to replace the Lagrangean of QCD at high energies. After including photons the Lagrangean can be even spontaneously broken down to low energies by allowing the scalar mesons to develop a complex-valued vacuum expectation value, which will provide the quarks with complex masses explaining their invisibility at low energies. Based on experience with the dynamical generation \cite{Delbourgo:1993dk,Deshpande:1984ke} of Linear Sigma Models we finally feel the need to perform a slight important modification of the value of the quartic coupling $\lambda$ having resulted from our simplistic ``mapping'' for a reason, which we will illustrate here on the basis of a much simpler ``Wess-Zumino-like'' Linear Sigma Model Lagrangean involving only one scalar $\sigma$, one pseudoscalar $\eta$, and $N_f$ Fermions $\psi$:
\begin{eqnarray} {\cal L} & = & \overline{\psi^c} \, \Big( \,\frac{i}{2} \!\! \stackrel{\;\,\leftrightarrow}{\partial\!\!\!/} \!\! -  M_\psi + g \, ( \sigma + i\, \eta \, \gamma_5 ) \Big) \, \psi \nonumber \\
 & & + \, \frac{1}{2} \, \big( (\partial \sigma)^2 - M^2_\sigma \,\sigma^2 \big)+ \frac{1}{2} \big( (\partial \eta)^2 - M^2_\eta \,\eta^2 \big) - \frac{\lambda}{4} \left( \sigma^2 + \eta^2 \right)^2 \; .
\end{eqnarray}
The quadratically divergent tadpole appearing in this theory can be absorbed by a suitable counterterm. As in the supersymmetric Wess-Zumino model \cite{Wess:1973kz} there can be extracted a relation between the Yukawa-coupling $g$ and the quartic coupling $\lambda$ by requiring an exact cancellation of quadratic divergencies on the level of $\sigma$- and $\eta$-selfenergies. A cancellation of quadratic divergencies up to two loops yields the important relation $g^2 N_f =  2\,\lambda -i\,\lambda^2 \,\frac{3}{4} \ln \left( \frac{4}{3}\right)$. The well known one loop relation \cite{Deshpande:1984ke} $g^2 N_f \simeq  2\,\lambda$ being obtained by skipping the term $O(\lambda^2)$ suggests, that a Linear Sigma Model with purely {\em imaginary} Yukawa coupling $g$ should be accompanied --- up to one loop --- by a {\em negative} quartic coupling $\lambda$. This explains partially the intimate relation between C.M.\ Bender's \cite{Bender:1998ke,Bender:1999ek} ``physical'' $i\,\phi^3$-theory and K.\ Symanzik's \& C.M.\ Bender's \cite{Bender:1999ek,Bender:1998ke}(See G.~'t Hooft \cite{Gross:1973id}!) ``physical'' $-\phi^4$-theory. Moreover suggests the (non-perturbative) two-loop relation between $g$ and $\lambda$ that at least one of these couplings have to be complex-valued for $\lambda$ finite \footnote{This might simultaneously explain why standard $\phi^4$-theory possesses a {\em non-trivial} phase when treated non-perturbatively \cite{Cea:2004ka}.}. A corresponding non-perturbative two-loop relation between $\lambda$ and $g$ for the Lagrangean Eq.\ (\ref{linsig1}) is in preparation. Recall, that it is (A)CQT allowing us to work {\em consistently} in the presence of {\em complex} masses and couplings! Due to these our observations we suggest to look for supersymmetric partners of standard model particles {\em not at high energies}, where we don't find them, yet {\em in the complex energy plane}, where they have escaped our observation by now! Similarly also the scalar Higgs-Boson may hide from present experiments not only due to Adler-like zeros of reaction amplitudes like in hadronic physics \cite{Bugg:2003kj}, yet also due to the feature of being a broad resonance whose pole is situated --- in correspondence to a complex-valued scalar vacuum condensate --- somewhere in the complex energy plane.
\section{Final Remarks}
We are fully aware that this introductory presentation and mentioned literature is far from being exhaustive. Many important and interesting issues and aspects of the comprehensive formalism have been omitted to keep the present size of the manusscript under control, yet may be certainly subject to and make part of a more detailed forthcoming outline. As an example we mention that the sketched complex probability concept going hand in hand with the presence of {\em non-Hermtitian density matrices} will have important implications not only to modern information theory and statistical physics, yet might also contribute e.g.\ to the resolution of certain quantum-theoretic paradoxa \cite{afriat1999}. Even though it should have become clear that renormalization in (A)CQT is performed --- even for fields with finite complex mass --- in complete analogy to traditional HQT, the (observable) consequences of imaginary parts of mass and coupling parameters to resulting finite quantities deserve certainly a more thorough investigation. The present discussion on PT-symmetric QM being a cousin of the Lee-model is reminding us again that we still have not made use of all possibilities which QT is offering to us. The existence of physically acceptable non-Hermitian Hamilton operators with a real spectrum opens interesting new fields in physics which have not been explored yet and which seem to allow simple solutions to tradionally complicated problems like e.g.\ asymptotic freedom and confinement in strong interactions having been unnecessarily constraint by the influential and impressive ``proofs'' e.g.\ by A.\ Zee, S.R.\ Coleman, D.R.\ Gross, F.\ Wilczek performed under too restraining assumptions. The formalism of (A)CQT presented here may be used conveniently to approach such problems and even problems with complex spectra without running into quantum-theoretic inconsistencies. As in the formalism of {\em Open Quantum Systems} (OQS) \cite{Kleefeld:Romano:2003ms,Kleefeld:Kossakowski2002}, where one needs to close the system by a ``heat bath'' (``reservoir'', ``environment'') before quantizing, we had to ``close'' a {\em causal} system in (A)CQT by adding the respective Hermitian conjugate {\em anticausal} system. To make (A)CQT causal, Poincar\'{e} invariant, analytic and local we had to apply the POSTULATE of {\em non-interaction} of the {\em causal} and the {\em anticausal} system. Such an uncorrelated situation between system and heat bath would be called in the formalism of OQS {\em complete positivity} \cite{Kleefeld:Romano:2003ms}. It is quite attractive to draw more correspondences between the formalisms of OQS and (A)CQT. As in Thermal Field Theory (TFT) \cite{Kleefeld:Weldon:1998yk}, but for different reasons and already at zero temperature, we observe also in (A)CQT a {\em doubling of degrees of freedom} due to the underlying indefinite metric. Differently from TFT or the respective Real Time Formulation (RTF) \cite{Sarkar:2000pg} the position of the vacuum at zero energy and the new antiparticle concept in (A)CQT does not allow to excite particle-antiparticle pairs by distribution functions having an overlap with negative energies due to finite temperature. Hence, in (A)CQT the excitation of particle-antiparticle pairs (even with complex masses) at finite temperature has to be resolved by distribution functions making use of positive energies only, requiring a modification of TFT and its RTF, which will simultaneously lead to the concept of a {\em causal heat bath}. A similar argument rules out Gribov's theory for confinement \cite{gribov2001} within (A)CQT.  

Our considerations showed that the way along which scientific progress proceeds is far from geodesic. P.A.M.\ Dirac tried to explain with his hole theory for electrons {\em protons} \cite{Dirac:1930,Quinn:2001kb}, while experimentalists disovered shortly afterwards \cite{Quinn:2001kb} the {\em positron}, i.e.\ the electron's antiparticle. Since then the positron has served as the example to explain the existence of Fermionic antiparticles on the basis of the negative energy states of the Dirac equation, while Dirac had parallelly developed for Bosons the orthogonal idea, that negative energy states are responsible for particle absorption. Now, after investigating the implications of causality to the antiparticle concept even for complex mass particles we had to notice, that even for Fermions negative energy states are responsible for annihilating positive energy particles and their positive energy antiparticles, and not for their existence. Surprisingly, in the resulting antiparticle concept Fermions and anti-Fermions being isospin partner show up to have --- without harm and contrary to the ``proof'' of V.B.\ Berestetskii \cite{ber1980a} --- the {\em same} intrinsic parity, like traditionally Bosons and anti-Bosons! 
Even though the original ingenious construction of Dirac is nearly replaced, the ideas of Dirac persist through the reveiled existence of antiparticles and a lot of important insights in the theoretical formalism of QT and particle physics. 
Unfortunately it seems that very influential still active authors of important ideas related to indefinite metric, negative norm states and complex ghosts are still denying the possibility of a meaningful use of complex poles in QT \footnote{E.G.C.\ Sudarshan abandons the most interesting part on {\em negative norm states} of a preprint \cite{sudarshan2003} with the title ``Non-relativistic proofs of the spin-statistics connection'' by commenting {\em ``$\ldots$ Stipulating that such negative norm states (negative Hilbert space metric!) cannot be present in any physical theory will eliminate the possibility of obtaining an inverted spin-statistics connection $\dots$''} and shifting some of the discussion on non-Hermtitian fields to an appendix, even if we have demonstrated above, that causal states being non-Hermitian necessarily consist of underlying normal {\em and} abnormal shadow states; N.\ Nakanishi still defending his discouraging conclusions with respect to his Complex-Ghost Relativistic Field Theory states: {\em ``$\ldots$ Since the
appearance of negative probability is the greatest problem in the indefinite
metric theory and since, historically, many great physicists proposed wrong
resolution of it, I cannot believe that you could resolve this problem
correctly. But since it seems to me that you understand the
indefinite-metric theory quite differently from the standard one,  it is
very difficult to discuss this problem with you. $\ldots$''} (N.~Nakanishi, private communication, 31.10.2003).}. Anyway --- young scientists and physics students should be obliged to read their most important and ``thrilling'' papers!

This work has been supported by the
{\em Funda\c{c}\~{a}o para a Ci\^{e}ncia e a Tecnologia} \/(FCT) of the {\em Minist\'{e}rio da Ci\^{e}ncia e da Tecnologia (e do Ensinio Superior)} \/of Portugal, under Grants no.\ PRAXIS
XXI/BPD/20186/99, SFRH/BDP/9480/2002, and POCTI/\-FNU/\-49555/\-2002.

\end {document}